\documentclass[twocolumn,showpacs,preprintnumbers,amsmath,amssymb,
 aps,
 prb,
 lengthcheck,%
]{revtex4-1}
\usepackage{graphicx}
\usepackage[english]{babel} 
\usepackage[utf8]{inputenc}
\usepackage{hyperref}
\usepackage{array}
\usepackage{xcolor}

\makeindex

\begin{document}
 
\title{Scattering Theory of Non-Equilibrium Noise and \\Delta $T$ current fluctuations through a quantum dot}

\author{A. Popoff,$^{1,2}$  J. Rech,$^1$ T. Jonckheere,$^1$ L. Raymond,$^1$ B. Grémaud,$^1$ S. Malherbe,$^{1,3}$ and T. Martin$^1$}
\affiliation{$^1$ Aix Marseille Univ, Université de Toulon, CNRS, CPT, IPhU, AMUtech, Marseille, France}
\affiliation{$^{2}$ Collège Tinomana Ebb de Teva I Uta, BP 15001 - 98726 Mataiea, Tahiti, French Polynesia}
\affiliation{$^{3}$ Département de Physique, Ecole Normale Supérieure, 45 Rue d'Ulm, 75005 Paris, France}
\date{\today}

\begin{abstract}
We consider the non-equilibrium zero frequency noise generated by a temperature gradient applied on a device composed of two normal leads separated by a quantum dot. We recall the derivation of the scattering theory for non-equilibrium noise for a general situation where both a bias voltage and a temperature gradient can coexist and put it in a historical perspective. 
We provide a microscopic derivation of  zero frequency noise through a quantum dot based on a tight binding Hamiltonian, which constitutes a generalization of the pioneering work of Caroli {\it et al.}\cite{caroli71} for the current  obtained  in the context of the Keldysh formalism. 
For a single level quantum dot, the obtained transmission coefficient entering the scattering formula for the non-equilibrium noise corresponds to a Breit-Wigner resonance.
We compute the delta-$T$ noise as a function of the dot level position, and of the dot level width, in the Breit-Wigner case, for two relevant situations which were considered recently in two separate experiments.\cite{lumbroso18,larocque20}  In the regime where the two reservoir temperatures are comparable, our gradient expansion shows that the delta-$T$ noise is dominated by its quadratic contribution, and is minimal close to resonance. In the opposite regime where one reservoir is much colder, the gradient expansion fails and we find the noise to be typically linear in temperature before saturating. In both situations, we conclude with a short discussion of the case where both a voltage bias and a temperature gradient are present, in order to address the potential competition with thermoelectric effects. 
 \end{abstract} 


\maketitle

\section{Introduction} 

Over the last four decades, theoreticians and experimentalists alike have focused  on the study  of non-equilibrium noise in quantum mechanically coherent mesoscopic and nanoscopic devices. By now, it is well recognized that non-equilibrium noise brings further information compared to a current diagnosis. For instance, non-equilibrium noise depends explicitly on the fermionic, bosonic, or even anyonic  statistics of the charge carriers which flow through the device. Moreover, when one is dealing with a system bearing strong correlations such as one dimensional wires or edge states of the Fractional Quantum Hall effect (FQHE), the monitoring of the Fano factor (the ratio between the zero frequency noise and the current) allows to identify the anomalous charges which are transmitted through the system. Typically, the non-equilibrium situation is achieved by imposing a bias voltage. Reviews on this topic\cite{blanter00,martin05} cover a broad range of nanoscopic devices, ranging from ballistic conductors, setups containing a resonant level, diffusive conductors, hybrid superconducting systems, one dimensional wires  where strong Coulomb interactions operate, chiral Luttinger liquids of the Fractional Quantum Hall effect,...  

Alternatively, one can in principle connect the sample to two reservoirs at different temperatures and work with a zero-bias voltage. While no net current is expected to flow in the device when electron-hole symmetry is respected, one expects in general (with or without electron-hole symmetry) a non-equilibrium noise signal which depends on the temperature gradient. This situation of  thermally induced non-equilibrium quantum mesoscopic/nanoscopic physics is nowadays referred to as ``caloritronics''.\cite{giazotto06}

\begin{figure}[tbp]
\centering
\includegraphics[width=0.48\textwidth]{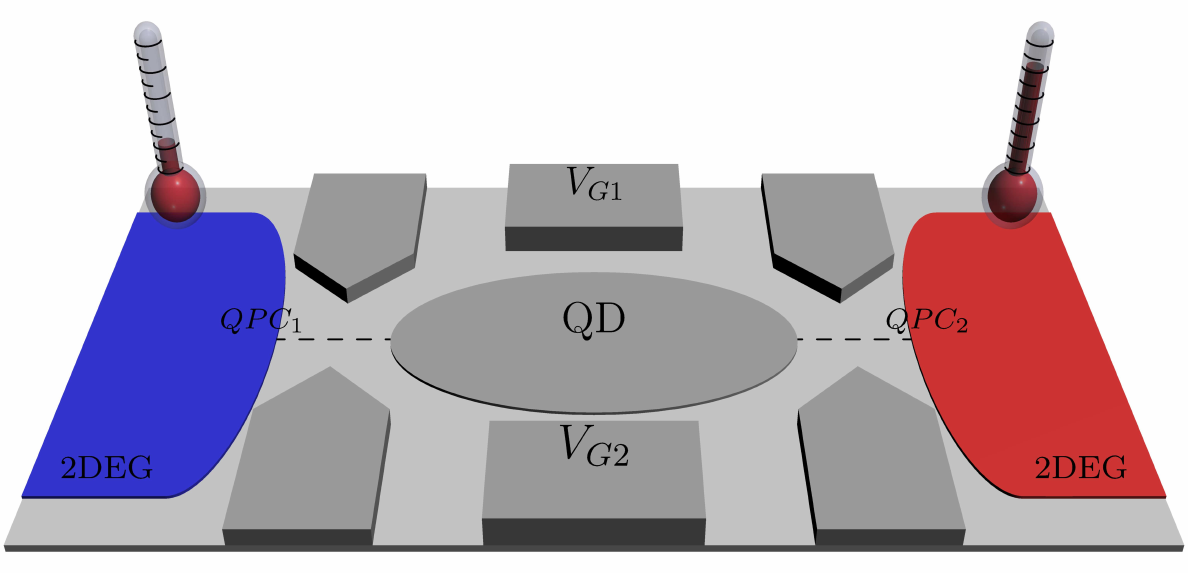}
\caption{Schematic description of a quantum dot connected to leads in a two dimensional electron gas: the quantum dot is defined by two quantum point contacts and side gates are applied to the dot.}\label{fig00}
\end{figure}

A recent experiment using atomic break junctions\cite{lumbroso18} - with an embedded hydrogen molecule between the leads - measured the noise signal, dubbed ``$\Delta T$ noise'' (DTN) in this situation. This generated a lot of interest for this previously undocumented source of non-equilibrium noise, both from the experimental \cite{sivre19,larocque20, duprez21} and the theoretical \cite{mu19, zhitlukhina20, rech20, hasegawa21, eriksson21} perspectives. Results from Ref.~\onlinecite{lumbroso18} were found to be in good agreement with the Scattering Theory of Non-Equilibrium Noise (STN) for a Fermi liquid (often referred to as the Landauer-B\"uttiker noise formula),
with the further assumption that the transmission coefficient $\tau$ is constant over characteristic energy/temperature ranges, which is well justified here as hydrogen bounds well with the metallic leads. A few theoretical works focused on going beyond this simple picture, by considering anomalous shot noise contributions, incorporating next order corrections\cite{mu19} or looking at generalized forms of DTN, considering charge or heat shot noise in the absence of charge or heat currents respectively.\cite{eriksson21}

The central purpose of the present work is to study DTN for a system where the transmission coefficient has a strong dependence in energy, by considering two normal metal leads (N) which are separated by a quantum dot (QD), as depicted in Fig.~\ref{fig00}. In this situation, the transmission coefficient is peaked around the location of the QD levels, which can be controlled experimentally with a gate voltage, while the peak width depends on the coupling strength between the QD and the leads. First, we find that when the temperatures of the two reservoirs are comparable, for a fixed average temperature, the total noise may {\it decrease} when increasing the temperature gradient, as a consequence of the thermal noise contribution being dominant. This encouraged us to argue that the definition of DTN as proposed in earlier works\cite{lumbroso18} has to be modified. We thus come up with an alternative definition, along the lines of the one proposed in Ref.~\onlinecite{hasegawa21}, that allows us to isolate the shot-noise like contribution (which always has a fixed, positive sign for fermions) independently of the energy profile of the transmission coefficient. Second, when the two reservoir temperatures differ substantially,\cite{larocque20} the gradient expansion fails. We address both points in this work and show that in the absence of voltage, depending on the chosen regime, DTN can lead to linear, quadratic, cubic (or combinations of either even odd powers) behavior as a function of $\Delta T$, or even to a saturation to a finite value. Note that Refs.~\onlinecite{mu19} and \onlinecite{eriksson21} studied somewhat similar setups in some specific limits, and our results match with their conclusions in the  regime of parameters where there is overlap with our investigation. We also consider the combination of both a voltage bias and a temperature gradient applied to the QD, in order to address the potential competition with thermoelectric effects. 

Along the way to characterize DTN for a QD embedded between normal metal leads, we discuss the origin of the STN formula. Furthermore, we provide a microscopic derivation of this formula using the same tight-binding Hamiltonian which was employed by Caroli {\it et al.}\cite{caroli71} to derive the current using the Keldysh formalism, from which a Breit-Wigner resonance transmission is obtained for a single level quantum dot. While this may look like an academic exercise, we believe that this derivation has not been published so far, and it has the advantage to be very general, valid for an arbitrary (multilevel) QD without having to define its shape/structure, and can even be extended to describe bosonic charge carriers.  

The paper is organized as follows. 
In Sec.~\ref{history}, we put the STN formalism in an historical perspective. Sec.~\ref{sec:microscopic} is devoted to the microscopic derivation of the STN formula. 
In Sec.~\ref{review}, we define the basic quantities needed to study DTN, and recall existing results for the case of a constant transmission. 
Sec.~\ref{energy} gives the general definitions of the important quantities for DTN in the case of an arbitrary energy dependent transmission coefficient. These definitions are then used in Sec.~\ref{breit_comparable} when the reservoir temperatures are comparable, and in Sec.~\ref{breit_cold} in the limit of a very cold reservoir, in order to obtain results for the case of a single level QD described by a Breit-Wigner resonance. We conclude in
  Sec.~\ref{conclusion}.  Throughout this work, we consider units in which Planck and Boltzmann constants are unity, \textit{i.e.} $\hbar=1$, $k_B=1$. Also, we focus on spinless particles, the addition of spin degrees of freedom (for spin $1/2$ electrons) yielding an overall factor $2$ on the current and noise in the absence of spin flip processes.   

\section{Historical perspective}
\label{history}

In the context of electronic quantum transport, it is quite difficult to identify the pioneers who were the first to derive non-equilibrium noise formulae involving a conductance, a resistance or a transmission probability. The story begins with the evolution of the Landauer conductance formula.

\subsection{Scattering theory pioneers}

Landauer initiated his effort on coherent electronic transport way ahead  of his time.\cite{landauer57} At that time, nanoscale devices were not available experimentally. His idea was that if the length of a device connected to Fermi liquid leads is sufficiently small (in fact smaller than  the quantum mechanical coherence length), the conductance $G$ should have a quantum mechanical nature. Assuming that the transmission coefficient $\tau$ characterizing the device is independent of energy, the Landauer conductance reads nowadays 
\begin{equation}
G=\frac{e^2}{2\pi}\tau ~.
\label{conductance}
\end{equation}
Note that the initial proposal of Landauer contained an (incorrect) extra factor $(1-\tau)^{-1}$ (see discussion below).
Yet quantum mechanics is based on a Hamiltonian formalism, which a priori does not  include dissipation, while the resistance $G^{-1}$, according to the (classical) Drude model, calls precisely for dissipation. This proposal was therefore totally counterintuitive, and it took some effort from Landauer to have his conductance formula recognized. 

After Anderson proposed his theory of localization,\cite{anderson58} and obtained the  Nobel prize for it, Landauer tried to argue more about his conductance formula, including as an aside his own approach to localization in one dimension.\cite{landauer70} The conductance formula was nevertheless still not fully accepted by the community at that time. 

The same year, Caroli {\it et al.}\cite{caroli71} computed the current for a one-dimensional tight binding model containing a central dot connected to leads, therefore achieving the first microscopic derivation of the energy dependent Landauer formula reading
\begin{equation}
\langle \hat{I} \rangle = \frac{e}{2 \pi} \int d\omega~ \tau  (\omega) \left[ f_1 (\omega) - f_2 (\omega) \right]~,
\label{landauerenergy}
\end{equation}
where $f_1$ and $f_2$ are the distribution functions in the two leads.

It took the work of Ref.~\onlinecite{anderson80} to rederive the Landauer formula and from then on it got accepted by the community. Landauer and coworkers subsequently published their seminal article for the multichannel noise formula,\cite{buttiker85} with an application to the Aharonov-Bohm effect when the conductor contains a small metallic ring. This work, among others, resolved the dilemma whether the conductance should be proportional to $\tau/(1-\tau)$ or $\tau$: the authors understood that the voltage had to be measured deep in the leads rather than in the direct vicinity of the scatterer. This implied that inelastic processes for outgoing electrons were essential to guarantee the presence of a Fermi sea in such (macroscopic) leads. This resolved the puzzle of why dissipation did not seem to appear in the Landauer conductance formula: dissipation is essential for thermal equilibration in the leads, which can then be described within the grand canonical ensemble. By that time, a majority of the experimental groups working in mesoscopic physics were using this formalism to interpret their results, and very successfully so.  

In the eighties, the connection between  the Landauer formula and the Green's functions of the device was approached via linear response theory by several authors,\cite{economou81,fisher81,baranger89} including generalizations to multichannel, multilead systems.

B\"uttiker used scattering theory to model four-terminal devices towards Onsager relations\cite{buttiker86} and also in order to make predictions on the integer quantum Hall effect\cite{buttiker88}.

Pioneering works on noise appeared in the mid eighties, first for tunnel junctions\cite{kulik84} and then in a weak link between two normal metals or superconductors.\cite{khlus87} The latter reference is rather technical, typically using the quasi classical Green’s function approach (Eilenberger equations) in specific systems described microscopically.

They were soon followed by contributions which used the scattering approach. Landauer proposed the wave packet approach, a first-quantized formulation, to compute the equilibrium noise and checked that the Johnson-Nyquist formula\cite{nyquist28}  was recovered. He found a match with the classical version of the theory using a Maxwell distribution. This approach described the incoming beam of electrons as a train of occupied/unoccupied states belonging to   orthogonal wave packets (which could overlap in space/time), restricted to a narrow energy range. The Pauli principle forbade that two incoming electrons could end up in the same wave packet state at either output. 

The second-quantized approach to noise, initiated by Lesovik, appeared around the same time.\cite{lesovik89,yurke90} There, the scattering matrix elements enter the expression for the current operator, and thermal equilibrium averages in the reservoirs are computed using Wick's theorem. Lesovik underestimated his own work, as it applies really to an arbitrary junction in the coherent regime containing potentially a scattering region between the leads encompassing the nano-devices. Both contain the celebrated proportionality $\tau(1-\tau)$ of the shot noise at zero temperature. Ref.~\onlinecite{buttiker90} reproduced this result and generalized it using linear algebra to multichannel, multiterminal mesoscopic devices.

\subsection{The wave packet approach}
\label{wave}

Ref.~\onlinecite{landauer91,martin92} derived non-equilibrium noise from the wave packet approach and generalized it to multichannel/multiterminal systems, with applications to a Hanbury-Brown and Twiss interferometer. In particular, the crossover between thermal and quantum shot noise was derived there. This approach is reviewed in the textbook Ref.~\onlinecite{imry97}. 

The wave packet approach\cite{landauer89,landauer91,martin92} is likely to be the most intuitive way to derive the STN formula. The current is viewed as a train of orthogonal wave packets (occupied by an electron or not), separated in time by an amount $h/\delta E$ (the wave packet has energies within the interval $[E,E+\delta E]$) in order to ensure orthogonality of successive wave packets  impinging from both sides of the scattering region. The zero frequency noise corresponds to an integral of the variance of the stochastic variable $g$ which counts whether a net electron charge has been transmitted on the right ($g=1$), on the left ($g=-1$) or no net electron charge has been transmitted ($g=0$): 
\begin{equation}
{\mathcal S}=\frac{e^2}{\pi}\int d\omega \left(\left\langle g^2 \right\rangle - \left\langle g \right\rangle^2\right)~.
\label{noisevariance}
\end{equation}

The only two possible events where a net charge $g=\pm 1$ is transmitted bear a probability $\tau f_{1,2} (1-f_{2,1})$: an electron from the left (right) is transmitted while no electron is incident from the right (left). Inserting this in Eq.~\eqref{noisevariance} leads directly to the STN formula 
 \begin{align}
 {\mathcal S}
= \frac{e^2}{\pi}  & \int d \omega 
 \left\{ \tau^2
 \left[ f_1 \left( 1 - f_1 \right) + f_2 \left( 1 - f_2 \right) \right]  \right. \nonumber \\
 & \left.
+ \tau \left( 1 - \tau \right)
 \left[ f_1 \left( 1 - f_2 \right) + f_2 \left( 1 - f_1 \right) \right]  
 \right\}.
 \label{eq:STNformula}
\end{align}
In Appendix~\ref{sec:wavepacket_appendix}, we give details about the derivation of the wave packet approach, with special emphasis on two-electron collisions (where the statistical weight for occupancy is $f_1f_2$) and where no contributions to the noise are generated. For illustrative purposes, in Appendix \ref{sec:wavepacket_appendix} we also illustrate the case of two boson collisions where bunching effects are manifest.

\section{Microscopic formulation}
\label{sec:microscopic}

\begin{figure}[tbp]
\centering
\includegraphics[width=0.48\textwidth]{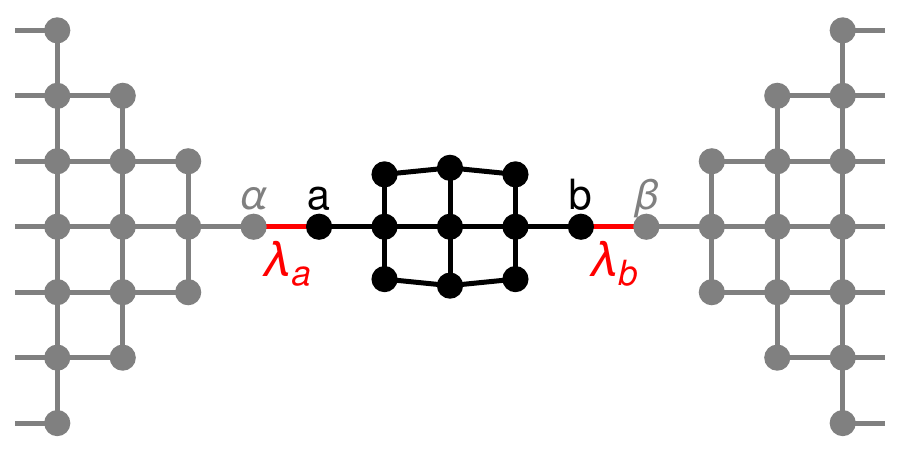}
\caption{Schematic representation of a metallic island connected to normal leads. In the text, we isolate the sites $\alpha$ and $\beta$ from the leads, and the sites $a$ and $b$ from the central metallic island. $\lambda_{a}$ (resp. $\lambda_b$) is the coupling constant between sites $a$ and $\alpha$ (resp. $b$ and $\beta$).}\label{fig:island}
\end{figure}

The present formulation is very general and applies equally to a single site dot and to a spatially extended island. Although in the following sections we are primarily interested in the case of a QD with a single energy level, for generality purposes, and in order to make the resulting expression useful in other contexts, we choose to present the derivation for the case of a metallic island with potentially many energy levels. The latter is connected to two normal metal leads, as schematically depicted in Figs.~\ref{fig00} and \ref{fig:island}. We compute the zero frequency noise using the Keldysh formalism applied to a tight binding model, in order to recover the STN formula. This constitutes an extension of the microscopic derivation of the Landauer formula proposed in the seventies.\cite{caroli71} 

\subsection{Model and Dyson equations}
\label{dyson}

We use the same formalism as in Refs. \onlinecite{cuevas96,cuevas99}, which considered current and noise in an atomic point contact. Greek (roman) indices are used for the lead (dot) degrees of freedom in this section.
The total Hamiltonian of the system can be decomposed into a contribution from the leads, one from the island and a final one describing the tunneling between the island and the leads: 
\begin{equation}
H=\sum_{j}H_j+H_D+H_T(t) ~ .
\end{equation}

Our starting point is the tunneling Hamiltonian between leads and central island of the form
\begin{equation}
H_T = \lambda_a \left( \psi_\alpha^\dagger \psi_a + \text{H.c.} \right) + \lambda_b \left( \psi_\beta^\dagger \psi_b + \text{H.c.} \right)~,
\end{equation}
where $\psi^\dagger_j$ is the creation operator for an electron on site $j$ while $\lambda_{a}$ (resp. $\lambda_b$) is the coupling constant between sites $a$ and $\alpha$ (resp. $b$ and $\beta$) as depicted in Fig.~\ref{fig:island}.
This Hamiltonian allows us to focus on the 4 sites $\alpha, a, b, \beta$. Note that the only assumption here is that there exists only a single injection point from each lead to the island. At this time, all microscopic details are kept arbitrary, and we do no need to specify the general Hamiltonian for the dot and leads. All this information is contained in the expression of the bare Green's functions. We therefore focus on the Green's functions constructed in this $4 \times 4$ site basis, introducing the bare Green's functions $\tilde{g}^{R/A}$ and tunneling matrix $\tilde{W}$ defined respectively as
\begin{equation}
\tilde{g}^{R/A} = 
\begin{pmatrix}
g_\alpha^{R/A} & 0 & 0 & 0 \\
0 & g_{aa}^{R/A} & g_{ab}^{R/A} & 0 \\
0 & g_{ba}^{R/A} & g_{bb}^{R/A} & 0 \\
0 & 0 & 0 & g_\beta^{R/A} 
\end{pmatrix}~,
\end{equation}
and
\begin{equation}
\tilde{W} = 
\begin{pmatrix}
0 & \lambda_a & 0 & 0 \\
\lambda_a & 0 & 0 & 0 \\
0 & 0 & 0 & \lambda_b \\
0 & 0 & \lambda_b & 0 
\end{pmatrix}~.
\end{equation}

The full Keldysh Green's functions are then defined as
\begin{equation}
G_{\mu \nu}^{\eta \eta'} (t, t') = - i \left\langle T_K \psi_\mu \left(t^\eta\right) \psi_\nu^\dagger \left({t'}^{\eta'}\right) \right\rangle~,
\end{equation}
where the Keldysh time indices $\eta, \eta'$ can be either $+$ or $-$ and the island/lead indices $\mu, \nu$ can be $\alpha, a, b$ or $\beta$.

Following the standard formalism, this then allows to define the retarded and advanced full Green's functions which satisfy the set of Dyson equations given by (omitting indices)
\begin{equation}
\tilde{G}^{R/A} = \tilde{g}^{R/A} + \tilde{g}^{R/A} \tilde{W} \tilde{G}^{R/A}~.
\label{eq:DysonRA}
\end{equation}
Similarly, for the $+-$ and $-+$ Keldysh components, they read
\begin{equation}
\tilde{G}^{\pm \mp} = \left( \tilde{1} + \tilde{G}^R \tilde{W} \right) \tilde{g}^{\pm \mp} \left( \tilde{1} + \tilde{W} \tilde{G}^A \right)~,
\label{eq:Gpmmp}
\end{equation}
where $\tilde{1}$ is the identity matrix in the site basis.

In Appendix \ref{app:micro}, we derive specific expressions for the Green's functions which are relevant for the calculation of the current and the noise.


\subsection{Current}

The current operator through site $\alpha$ can be written as
\begin{equation}
I_\alpha = i e \lambda_a \left( \psi_\alpha^\dagger \psi_a - \psi_a^\dagger \psi_\alpha \right)~.
\end{equation}

Taking then the thermodynamic average, the current reads
\begin{equation}
\langle I_\alpha \rangle= - e \lambda_a \int \frac{d\omega}{2 \pi}  \left[ G_{\alpha a}^{+-} (\omega) - G_{a \alpha}^{+-} (\omega) \right]~.
\end{equation}

At this stage, it is useful to recall that the $\pm \mp$ Keldysh components of the bare Green's functions for the leads are conveniently expressed in terms of the density of states $\rho_j$ ($j=\alpha, \beta$) and the occupation as
\begin{align}
g_{j}^{+-} =& \pm2 i \pi \rho_j f_j ~,\\
g_{j}^{-+} =& \mp 2 i \pi \rho_j \left( 1 \mp f_j \right)~,
\end{align}
where the upper (lower) sign applies for fermions (bosons). 
It turns out that these are the only two bare Green's functions of the system which enter the expression for the current and noise. Theoretically, contributions involving the central island $\pm \mp$ bare Green's functions can be shown to involve the same infinitesimal parameter which enters the lead advanced and retarded Green's functions\cite{rechprivate} so that they give no contribution to transport. This point was overlooked previously,\cite{caroli71} where it was assumed that the central island Green's function had a zero density of states in the bias window, limiting therefore the discussion to the case of an insulating barrier. 
 
Working out the Keldysh Green's functions explictily (see Appendix \ref{app:micro} for details), one is finally left with
\begin{equation}
\langle I_\alpha \rangle = e  \int \frac{d\omega}{2 \pi} \left( 2 \pi \right)^2  \lambda_a^2 \lambda_b^2 \rho_\alpha \rho_\beta G_{a b}^R  G_{b a}^A \left(  f_\alpha - f_\beta \right)~,
\label{eq:microcurrentformula}
\end{equation}
where we recover the energy dependent Landauer formula Eq.~\eqref{landauerenergy}, where the transmission reads:
\begin{equation}
\tau(\omega) = \left( 2 \pi \right)^2  \lambda_a^2 \lambda_b^2 \rho_\alpha(\omega) \rho_\beta(\omega) G_{a b}^R(\omega)  G_{b a}^A(\omega)~.
\label{eq:Tofomega} 
\end{equation}

In the event that the QD contains only a single energy level $H_D=\epsilon \psi^\dagger_a\psi_a$, $a=b$ in the above equation, and in the wide band limit, one recovers the transmission coefficient corresponding to a Breit-Wigner resonance:
\begin{equation}
\tau(\omega)=\frac{(\Gamma_1+\Gamma_2)^2}{(\omega-\epsilon)^2+(\Gamma_1+\Gamma_2)^2}~,
\label{eq:defBW}
\end{equation}
with the tunneling rates $\Gamma_j= 2\pi \rho_j \lambda_a^2$. 
 
\subsection{Noise}

Introducing the current deviation $\delta I_\mu (t)=I_\mu (t)-\langle I_\mu\rangle$, the real time current correlator is expressed, using Wick's theorem, as: 
\begin{align}
{\mathcal S}_{\mu \nu} (t,t') 
=& \langle T_K  \delta I_\mu (t^-) \delta I_\nu (t'^+) \rangle \nonumber \\
=& - e^2 \lambda_m \lambda_n \left[  G_{\mu n}^{-+} (t,t') G_{\nu m}^{+-} (t',t) \right.  \nonumber \\
 &  -  G_{\mu \nu}^{- +} (t,t') G_{n m}^{+-} (t',t) -  G_{m n}^{-+} (t,t') G_{\nu \mu}^{+ -} (t',t) \nonumber \\
 & \left. + G_{m \nu}^{- +} (t,t') G_{n \mu}^{+ -} (t',t)  \right],
\end{align}
where we use compact notations implying that $\mu = \alpha \Rightarrow m=a$ and $\mu = \beta \Rightarrow m=b$ (and similarly for $\nu$ and $n$).

The zero frequency noise evaluated at lead $\alpha$ (the final result is identical for lead $\beta$ as expected) is given by:
\begin{align}
 {\mathcal S} = &- e^2 \lambda_a^2 \int \frac{d \omega}{2 \pi} \left[ G_{\alpha a}^{-+} (\omega) G_{\alpha a}^{+-} (\omega) -  G_{\alpha \alpha}^{- +} (\omega) G_{a a}^{+-} (\omega)\right.\nonumber\\
 & \qquad  -  \left. G_{a a}^{-+} (\omega) G_{\alpha \alpha}^{+ -} (\omega) +  G_{a \alpha}^{- +} (\omega) G_{a \alpha}^{+ -} (\omega) \right].
 \label{eq:Saadef}
\end{align}

The integrand of this expression is computed in Appendix \ref{app:micro}, with the following result for the noise:
 \begin{widetext}
 \begin{align}
 {\mathcal S} &= e^2 \lambda_a^2 \int \frac{d \omega}{2 \pi} 
\left\{
\lambda_a^2 \lambda_b^4 \left( G_{ab}^R G_{ba}^A \right)^2 \left(2 \pi \right)^4 \rho_\alpha^2 \rho_\beta^2
\left[ f_\alpha \left( 1 - f_\alpha \right) + f_\beta \left( 1 - f_\beta \right) \right] \right. \nonumber\\
&\left.~~~~~~~~~~~~~~~~~~~+ \lambda_b^2 G_{ab}^R G_{ba}^A \left(2 \pi \right)^2 \rho_\alpha \rho_\beta  \left[ 1 - \left( 2 \pi \right)^2 \lambda_a^2 \lambda_b^2 G_{ab}^R G_{ba}^A \rho_\alpha \rho_\beta  \right] 
\left[ f_\alpha \left( 1 - f_\beta \right) + f_\beta \left( 1 - f_\alpha \right) \right]  
\right\}  
\nonumber \\
&= \frac{e^2}{\pi}  \int d \omega 
 \left\{ \tau(\omega)^2
 \left[ f_\alpha \left( 1 \mp f_\alpha \right) + f_\beta \left( 1 \mp f_\beta \right) \right]  
+ \tau(\omega) \left( 1 - \tau(\omega) \right)
 \left[ f_\alpha \left( 1 \mp f_\beta \right) + f_\beta \left( 1 \mp f_\alpha \right) \right]  
 \right\}~,
 \label{eq:micronoise}
\end{align}
 \end{widetext}
where we recover the STN formula of Eq.~\eqref{eq:STNformula} for fermions and bosons, with the same energy-dependent transmission as defined above for the current [see Eq.~\eqref{eq:Tofomega}].

\section{Basics of $\Delta T$ noise}
\label{review}

For a junction between two normal metal reservoirs with an arbitrary energy-dependent transmission
$\tau(\omega)$,  
the current is given by the energy dependent Landauer formula, Eq.~\eqref{landauerenergy}. 
In full generality, the normal metal reservoirs, labeled $j=1,2$ respectively, are characterized by their chemical potential $\mu_j$ and temperature $T_j$, leading to a description in terms of the Fermi distribution:
\begin{equation}
f_j (\omega) = \frac{1}{e^{(\omega - \mu_j)/T_j}+ 1}~.
\end{equation}
As we are concerned with DTN, in this section we consider the case where the two distribution
functions $f_1$ and $f_2$ differ only by their temperature ($T_1 \neq T_2$). 

The zero frequency noise is defined as:
\begin{equation}
\mathcal{S} = 2 \int dt \left[ \langle \hat{I} (t) \hat{I} (0) \rangle -  \langle \hat{I} (t) \rangle \langle \hat{I} (0) \rangle \right] ~.
\end{equation}
The STN formula reads in full generality: 
\begin{equation}
{\mathcal S}={\mathcal S}_1\ + {\mathcal S}_2 \; ,
\label{STN1}
\end{equation}
where we chose to isolate two separate contributions
\begin{align}
{\mathcal S}_1 &= \frac{e^2}{\pi} \int d\omega  ~ \tau  (\omega) \left\{ f_1 (\omega) \left[ 1 - f_1 (\omega) \right] \right. \nonumber \\
& \qquad \qquad \qquad \qquad \left. + f_2 (\omega) \left[ 1 - f_2 (\omega) \right] \right\} ~,
\label{STN2}\\
{\mathcal S}_2 &= \frac{e^2}{\pi} \int d\omega ~ \tau  (\omega) \left[ 1- \tau  (\omega) \right] \left[ f_1(\omega) - f_2(\omega) \right]^2 ~,
\label{STN3}
\end{align}
with ${\mathcal S}_{1,2}>0$. When the junction is biased, the contribution
${\mathcal S}_1$ corresponds to the transmission of the thermal noise emanating from each of the two reservoirs and is present even in an equilibrium situation, while ${\mathcal S}_2$ is  a purely non-equilibrium contribution leading to quantum shot noise. Note that here, we purposely adopt a different convention for the definition of ${\mathcal S}_1$ and ${\mathcal S}_2$ compared to the supplemental material of Ref.~\onlinecite{lumbroso18}.

For the rest of this section, we now concentrate on the case of a constant transmission with $\tau(\omega)=\tau$.
Let us first recall that, for a device at finite temperature $T$ and at equilibrium (no bias), the noise yields the Johnson-Nyquist form\cite{nyquist28}
\begin{equation}
{\mathcal S}= 4 T G ~,
\label{eq:JNnoise}
\end{equation}
with the conductance $G$ given by the Landauer formula, Eq.~\eqref{conductance}. The opposite limit of the noise in the presence of a bias $V$, at zero temperature, yields the quantum shot noise formula
\begin{equation}
{\mathcal S}= 2 \langle \hat{I} \rangle (1-\tau)=\frac{e^3}{\pi} \tau(1-\tau)V~.
\label{tminust}
\end{equation}

\subsection{Comparable reservoir temperatures}
\label{comparable}

For the case of a temperature gradient, in the regime where the temperature difference $\Delta T=T_1-T_2$ between the reservoirs is small compared to the average temperature $\bar{T}=(T_1+T_2)/2$ it is useful to expand the non-equilibrium noise in an even power series of $\Delta T/(2\bar{T})$:
\begin{equation}
{\mathcal S}= {\mathcal S}_0 \left[1+ C_2 \left(\frac{\Delta T}{2\bar{T}}\right)^2 + C_4 \left(\frac{\Delta T}{2\bar{T}}\right)^4 +...\right]~,
\end{equation} 
where ${\mathcal S}_0$ is the thermal equilibrium noise at temperature $\bar{T}$.
The coefficients of the series, $C_2$, $C_4$, ... are entirely due to the non-equilibrium contribution
$\mathcal{S}_2$, as $\mathcal{S}_1$ does not depend on $\Delta T$, and thus 
$\mathcal{S}_1 = \mathcal{S}_0=4  \bar{T} G$.
 The values of the coefficients are:
\begin{align}
   C_2 =&4(1-\tau)\int_{-\infty}^{+\infty}\frac{u^2}{\cosh^4(u)}du\nonumber\\
    =&(1-\tau)\left(\frac{\pi^2}{9}-\frac{2}{3}\right)~,\label{c2constant}\\
   C_4  =& -(1-\tau)\left(\frac{7\pi^4-75\pi^2+90}{675}\right)~.
   \label{c4constant}
\end{align}
In Ref.~\onlinecite{lumbroso18} the coefficient $C_2$ was extracted from experimental data and
successfully compared to theory.
This completes the summary of DTN at constant transmission, albeit in the situation where the temperatures of both reservoir are comparable, which means $\Delta T \ll \bar{T}$.

Note that in a recent work,\cite{rech20} we studied DTN in the strongly correlated regime of the Fractional Quantum Hall system, where in the weak backscattering situation, Laughlin quasiparticles - not electrons - tunnel from one edge state to the other at the location of a quantum point contact. These Laughlin quasiparticles are anyonic excitations, with both fractional charge and statistics.\cite{wilczek82} We argued in Ref.~\onlinecite{rech20} that one can obtain a signature of anyon statistics via the measurement of DTN. Indeed, we found that the $C_2$ and $C_4$ ($\vert C_2\vert \gg \vert C_4\vert$) coefficients are both {\it negative}. On the contrary, in the strong backscattering regime, where only electrons can tunnel between the two semi-infinite chiral Luttinger liquids - instead of anyons - we obtained $C_2>0$ ($\vert C_2\vert \gg \vert C_4\vert$) as for Fermi liquids.

\subsection{Cold reservoir case}
\label{cold}

An alternative point of view for the study of DTN is to consider the case where one of the reservoirs is placed at very low temperature, $T_2 \ll T_1$, so that the temperature bias is much larger than the temperature of the coldest lead (which, for all practical purposes, can then be taken to 0). The gradient expansion of the noise is not valid anymore as $\bar{T}\sim T_1/2$ and $\Delta T\sim T_1\equiv T$. This situation was studied both theoretically and experimentally\cite{larocque20} for the case of a normal metal tunnel junction ($\tau\ll 1$) which to a good approximation has a constant transmission coefficient when the metallic bandwidth is large compared to the temperature gradient. In this situation, the total noise reads
\begin{equation}
{\mathcal S}=4 \log{2} ~ T G~.
\label{reulet}
\end{equation}
Interestingly, although  the system is placed far from equilibrium, this resembles a Johnson-Nyquist dissipation-fluctuation result [see Eq.~\eqref{eq:JNnoise}] albeit with an additional factor $\log{2}$. The latter was attributed in Ref.~\onlinecite{larocque20} to the Landauer principle,\cite{landauer61} which states that when a bit of information is erased (here the information corresponds to the final outcome, whether the electron has been transmitted or reflected), dissipation - and therefore noise - is unavoidable. The authors suggested a possibly deeper connection with information theory, which sets the stage for dissipationless, reversible quantum computation.  Eq.~\eqref{reulet} was in very good agreement with the results of caloritronic experiments.

\section{$\Delta T$ noise for an energy dependent transmission}
\label{energy}

We now study  the case of a junction with an arbitrary energy-dependent transmission, detail the contributions $\mathcal{S}_1$ and $\mathcal{S}_2$, and introduce an alternative definition of DTN in this case, that turns out to be more adequate. The definitions introduced here are quite general and will be used in the next section when we consider specifically the case of a single level quantum dot.

\subsection{General remarks}

We thus consider an unbiased mesoscopic device connected to two leads.
The first (thermal-like) contribution to the noise [see Eq.~\eqref{STN2}] is readily rewritten as
\begin{equation}
{\mathcal S}_1 = - \frac{e^2}{\pi} \int d\omega  ~ \tau  (\omega) \left[ T_1 \partial_\omega f_1 (\omega) + T_2 \partial_\omega f_2 (\omega) \right]~.
\end{equation}
Keeping in mind that the chemical potentials are equal, the two distribution functions $f_{1,2}$ only differ by their temperature, allowing us to rewrite the expression for ${\mathcal S}_1$ as
\begin{equation}
{\mathcal S}_1 = 2 \left[ T_1 G_d(T_1) + T_2 G_d (T_2) \right]~,
\label{eq:S1Gd}
\end{equation}
where we introduced the zero bias differential conductance:
\begin{equation}
G_d (T) = \left. \frac{\partial \langle \hat{I} \rangle }{\partial V} \right|_{V=0} =  - \frac{e^2}{2 \pi} \int d\omega~ \tau  (\omega) \frac{\partial f (\omega) }{\partial \omega}~.
\label{eq:conductance}
\end{equation}

The expression for the second (shot noise like) contribution ${\mathcal S}_2$ cannot be further simplified compared to the definition of Eq.~\eqref{STN3}. In practice, this is the contribution we are mostly interested in.
An important difference of the energy dependent transmission case, compared to the constant transmission case, is the definition of the reference equilibrium noise. 
For the constant transmission case, one has $\mathcal{S}_0 = \mathcal{S}_1$, and thus
defining the excess noise as $\mathcal{S}-\mathcal{S}_0$ allows to keep only the shot-noise like contribution $\mathcal{S}_2$ and to remove the thermal-like contribution 
$\mathcal{S}_1$.  This, however, no longer works in the energy dependent transmission case, as $\mathcal{S}_0 \neq \mathcal{S}_1$. It follows that $\mathcal{S}-\mathcal{S}_0$ therefore contains contributions from both $\mathcal{S}_1$ and $\mathcal{S}_2$. One way to circumvent this is to use a \emph{different} definition of the excess noise:\cite{hasegawa21}
\begin{equation}
\Delta \mathcal{S} (T_1,T_2) = \mathcal{S}(T_1,T_2) - 
  \frac{1}{2} \left[ \mathcal{S}(T_1,T_1) + \mathcal{S}(T_2,T_2) \right]~.
\label{proper_noise_def}
\end{equation} 
One can easily check that, with this definition, $\Delta \mathcal{S}$ simply reduces 
to $\mathcal{S}_2$ for \emph{any} transmission $\tau(\omega)$, and thus contains only shot noise like contributions. In the case of an energy independent transmission, this definition is of course equivalent to the one used before. Experimentally, the measurement of this excess noise thus requires three separate noise measurements (two thermal noise measurements at different temperatures, followed by a temperature gradient induced non-equilibrium noise measurement). It enables, however, to bypass the need for performing measurements at the average temperature $\bar{T}$, which can be an advantage in some experimental devices where fine tuning of the temperature is difficult.

\begin{figure*}[tbp]
	\centering
		\includegraphics[height=0.31\textwidth]{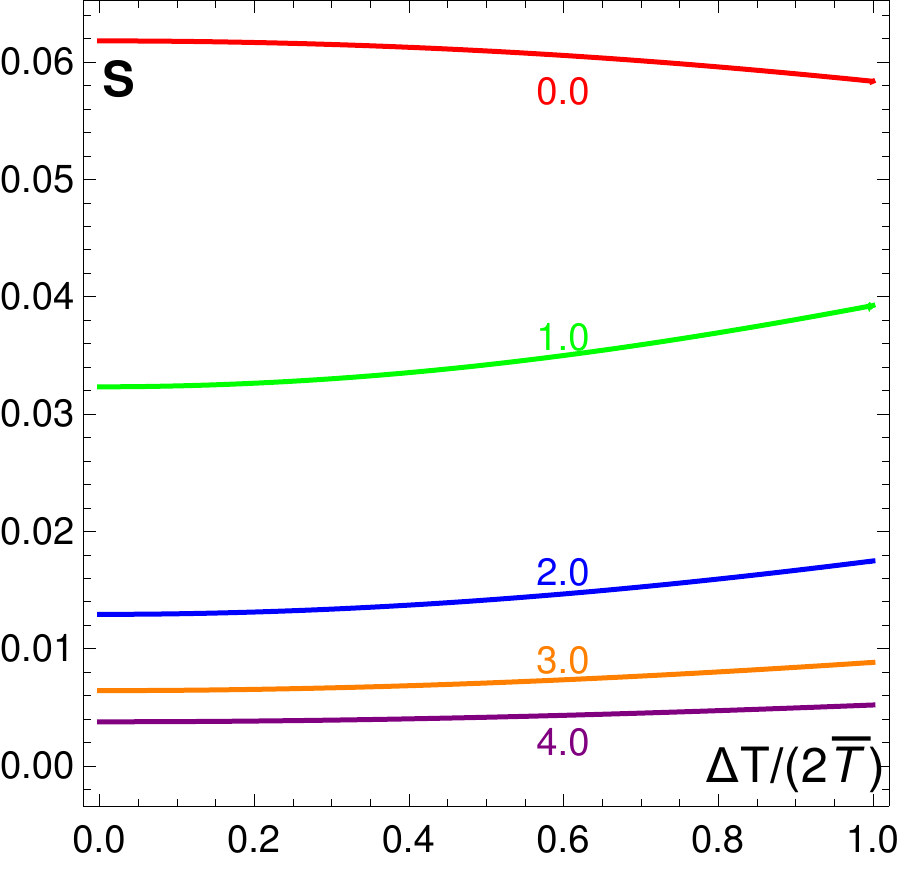}
		\includegraphics[height=0.31\textwidth]{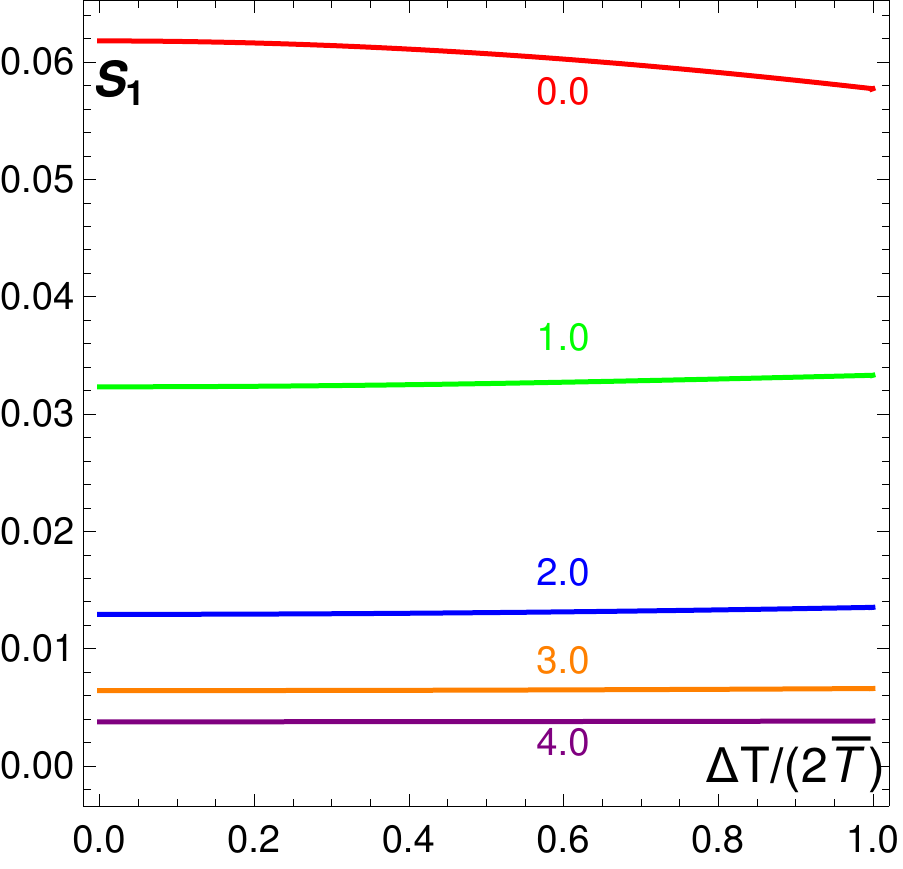}
		\includegraphics[height=0.31\textwidth]{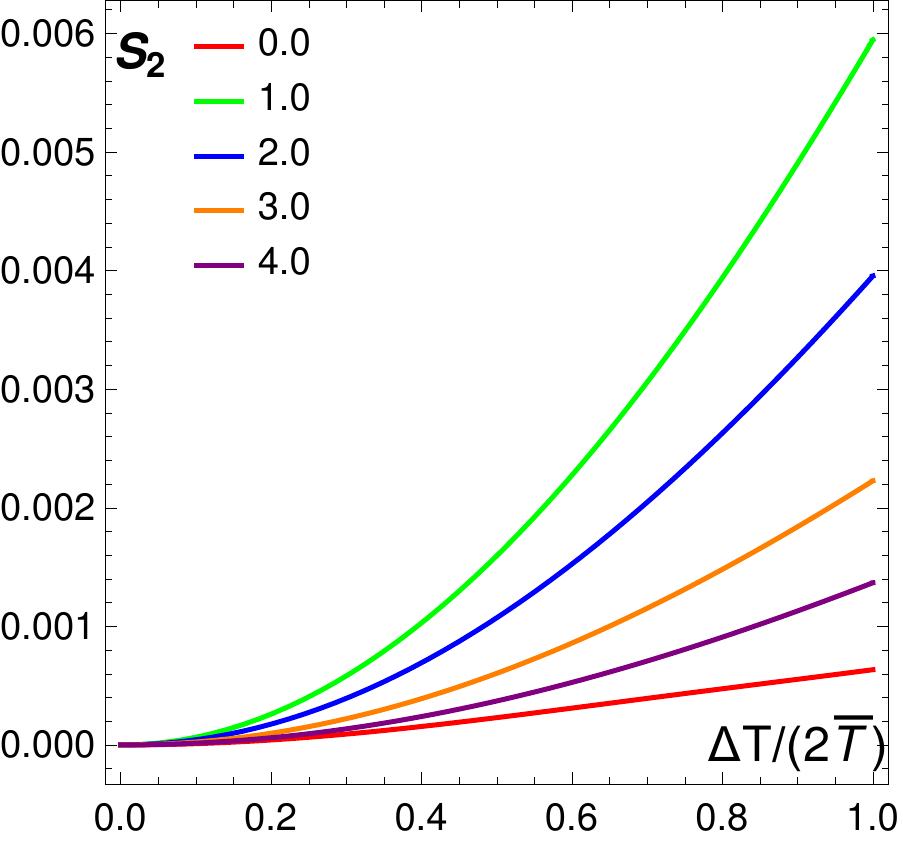}
		\caption{Full noise $\mathcal{S}(\Delta T)$ (left), thermal-like contribution $\mathcal{S}_1$ (center) and shot-noise-like contribution $\mathcal{S}_2$ (right) in units of $e^2 \Gamma$, as a function of $\Delta T/2\bar T$ for different values of the dot level $\epsilon/\Gamma=0,1.,2.,3.,4.$ at $\bar T=0.1\Gamma$.}
		\label{fig:Sdotfull}
\end{figure*}

\subsection{Comparable reservoir temperatures}

In the regime of small temperature gradient $\Delta T \ll \bar{T}$, ${\mathcal S}_1$ can be expanded in powers of $\Delta T$ directly from the temperature dependence of the conductance:
\begin{align}
{\mathcal S}_1 =& {\mathcal S}_0 + \left( \frac{\Delta T}{2} \right)^2 \left[ \partial_T^2 \left( 2 T G_d \right) \right]_{\bar{T}} \nonumber \\
 & \qquad +\frac{1}{6} \left( \frac{\Delta T}{2} \right)^4   \left[ 4\partial^3_T G_d+ T\partial^4_T G_d \right]_{\bar{T}}~,
  \label{gradient_power}
  \end{align}
with ${\mathcal S}_0 = 4 \bar{T} G_d(\bar{T})$. There we obtain a thermal noise, albeit with the Landauer conductance replaced by the differential conductance.
From this, one readily sees that the sign of the $\Delta T^2$ term of the $\mathcal{S}_1$ contribution to the noise is directly related to the way the conductance of the junction varies as a function of the mean temperature, and could very well be negative for a specifically designed transmission $\tau (\omega)$. 
In particular, assuming a power-law behavior for the conductance at low temperature, $G(T) \sim \alpha T^\gamma$, one is left with 
\begin{equation}
{\mathcal S}_1 = {\mathcal S}_0 \left[ 1 + \left( \frac{\Delta T}{2 \bar{T}} \right)^2 \frac{\gamma (\gamma +1)}{2} \right] ~.
\label{power_law}
\end{equation} 
For exponents $\gamma$ satisfying $-1 < \gamma < 0$, this means that
the total noise $\mathcal{S}$ can be decreasing when increasing $\Delta T$ (depending on the magnitude of the contribution from ${\mathcal S}_2$, which is always positive).\footnote{Note that in the FQHE case,\cite{rech20} the conductance is indeed a power-law with an exponent $\gamma = 2 (\nu-1)$ but negative delta-$T$ noise is obtained over a different interval than $1/2 < \nu < 1$.} 

Alternatively, one can disregard the connection to the zero bias differential conductance and expand the Fermi function as in the constant transmission case, yielding
\begin{align}
{\mathcal S}_1 = {\mathcal S}_0 - \frac{e^2}{\pi}  &  \int d\omega \tau(\omega)\left\{ \left( \frac{\Delta T}{2} \right)^2 \partial_\omega \left[ \partial^2_T \left( T f \right) \right]_{\bar{T}} \right. \nonumber \\
 & +  \left.\frac{1}{3} \left( \frac{\Delta T}{2} \right)^4   \partial_\omega \left[ \left( \partial^3_T f  \right)_{\bar{T}} + \frac{\bar{T}}{4}  \left( \partial^4_T f  \right)_{\bar{T}} \right] \right\} .
 \label{s1_gradient}
  \end{align}

This can be similarly carried out for the noise contribution ${\mathcal S}_2$, where keeping only terms up to $O \left( \Delta T^4 \right)$, one readily shows that
\begin{align}
{\mathcal S}_2 = \frac{e^2}{\pi} 4 \bar{T}^2 & \int d\omega ~ \tau  (\omega) \left[ 1- \tau  (\omega) \right]  
\left\{ \left( \frac{\Delta T}{2 \bar{T}} \right)^2 \left[ \partial_Tf \right]_{\bar{T}}^2\right.  \nonumber\\
& \left.+ \frac{\bar{T}^2}{3} \left( \frac{\Delta T}{2 \bar{T}} \right)^4 \left[  \left(\partial_Tf \right) \left( \partial^3_Tf \right) \right]_{\bar{T}}\right\}~.
\end{align}

There is not much one can do at this stage to further simplify the expressions for ${\mathcal S}_1$ and ${\mathcal S}_2$ without making further assumptions about the transmission function $\tau (\omega)$. Before fully tackling the case of a Breit-Wigner resonance, a first step to go beyond the constant transmission results is to include small deviations, performing an expansion near the Fermi energy, $\tau (\omega) \simeq \tau + \omega \tau'$. This was considered in a previous work~\cite{mu19} and we recover here similar results in the form
\begin{align}
{\mathcal S}_1 &= 4 \bar{T} G ~, \\
 {\mathcal S}_2 &= \left( \frac{\Delta T}{2 \bar{T}}\right)^2  4 G_0 \bar{T}  \left[   \tau (1-\tau)  \left(\frac{\pi^2}{9}-\frac{2}{3}\right) \right. \nonumber \\
 & \qquad \qquad \left.  - \left( \tau' \bar{T} \right)^2  \pi^2 \frac{7 \pi^2-60}{45} \right]~,
\end{align}
where $G_0 = \frac{e^2}{2 \pi}$ is the quantum of conductance.

Interestingly, when electron-hole symmetry is broken, that is when $\tau(\omega)$ is not an even function of energy anymore, the current is non-vanishing and  has an expansion in terms of odd powers of $\Delta T$:
\begin{equation}
\langle I\rangle = \sum_{n=0}D_n \left(\frac{\Delta T}{2 \bar{T}}\right)^n~,
\end{equation} 
where all even powers vanish when the two chemical potentials are equal, and in this situation we have:
\begin{align}
D_1  =& -\frac{e}{\pi}\int d\omega \tau(\omega) \omega\partial_\omega f~,
\\
D_3 =& -\frac{e}{\pi}\int d\omega \tau(\omega) \left\{ \omega\partial_\omega f+ \omega^2\partial^2_\omega f + \frac{1}{6}\omega^3\partial^3_\omega f \right\}~.
\end{align}

\subsection{Cold reservoir case}

In the cold reservoir regime, the two contributions to the noise can be written as 
\begin{align}
{\mathcal S}_1 &= 2 T G_d (T) ~, \\
{\mathcal S}_2 &= \frac{e^2}{\pi} \int d\omega ~ \tau  (\omega) \left[ 1- \tau  (\omega) \right] \left[ f (\omega) - \theta(-\omega) \right]^2 ~,
\end{align}
and no further simplification can be achieved without specifying further the shape of the transmission coefficient.

Again, one option is to resort to an expansion of the transmission beyond the constant case. While this leads to a rather trivial contribution for the thermal-like noise, ${\mathcal S}_1 = 2 T G$, the shot-noise like contribution reads
\begin{align}
{\mathcal S}_2 &= \frac{e^2}{\pi} \int d\omega ~ \left[  \tau \left( 1- \tau \right) -  \omega^2 \left( \tau' \right)^2  \right] \left[ f (\omega) - \theta(-\omega) \right]^2 \nonumber \\
&=2 G_0 T \left\{  \tau \left( 1- \tau \right) \left( 2 \log 2 -1 \right) \right. \nonumber \\
&  \qquad  \qquad \left.  -  \left( T \tau' \right)^2  \left[  3 \zeta(3) -2 \zeta(2)  \right]  \right\} ~,
\end{align}
where, in addition to the $\log 2$ term reminiscent of Ref.~\onlinecite{larocque20}, one readily sees that the first leading correction is always negative, and could be significant for high enough temperature $T$ of the hot reservoir.

\section{Breit-Wigner $\Delta T$ noise: comparable reservoir temperatures}
\label{breit_comparable}

In this section we consider a symmetric device composed of a single level QD in the absence of voltage bias, with reservoir temperatures $T_1\sim T_2$. We focus on symmetric coupling to the leads for simplicity. The transmission coefficient then corresponds to a Breit-Wigner resonance, Eq.~\eqref{eq:defBW}, with equal escape rates $\Gamma_1=\Gamma_2\equiv \Gamma/2$, namely
\begin{equation}
\tau (\omega) = \frac{\Gamma^2}{(\omega-\epsilon)^2+\Gamma^2}~.
\label{eq:BWformula}
\end{equation}
Our goal here is to study the behavior of the noise created by the temperature gradient  as a function of the parameters $\bar{T}/\Gamma$ and $\epsilon/\Gamma$. $\epsilon\neq 0$ corresponds to a breaking of electron-hole symmetry. The noise is computed
using the general equations Eqs.~\eqref{STN1}-\eqref{STN3}.

\subsection{Zero voltage case}

We choose $\Delta T/\bar{T} \ll 1$ and thus characterize the noise by studying the coefficients of the temperature gradient expansion ($C_2$, $C_4$, etc.) as in Sec.~\ref{comparable}. 
However, before considering the gradient series of the purely shot-noise-like contribution $\mathcal{S}_2$, it is relevant to plot the total noise of Eq.~\eqref{STN1}. Fig.~\ref{fig:Sdotfull} (left) shows
the full noise $\mathcal{S}$ as a function of the temperature bias $\Delta T$, for several values of $\epsilon/\Gamma$, and a mean temperature $\bar{T}=0.1 \Gamma$. The curves show immediately that, when the the QD level is on resonance ($\epsilon=0$), or close enough to resonance, the noise is a decreasing function of the temperature bias $\Delta T$. 

This result may seem counter-intuitive: for a given average temperature, the total noise on resonance is reduced when a larger temperature gradient is imposed! However, as explained in the previous section, this decreasing behavior is due to the thermal-like contribution to the noise $\mathcal{S}_1$. The  prefactor of $\Delta T^2$ in $\mathcal{S}_1-\mathcal{S}_0$ - [see Eqs.~\eqref{gradient_power} and \eqref{s1_gradient}] - is negative, and dominates in $\mathcal{S}$. This is attributed to the power law behavior (see Sec.~\ref{energy}) of the differential conductance $G_d(T)$ associated with the Breit-Wigner transparency $\tau(\omega)$.

It is therefore only for a strongly detuned dot ($\epsilon > \mbox{several } \Gamma$) that the noise becomes an increasing function of $\Delta T$. 

This fact is further illustrated in Fig.~\ref{fig:Sdotfull} (center) and Fig.~\ref{fig:Sdotfull} (right), where the separate contributions $\mathcal{S}_1$ and $\mathcal{S}_2$ are shown.
One can see that $\mathcal{S}_2$ has a  $\Delta T^2$ increasing contribution, while the $\mathcal{S}_1$ (also quadratic in $\Delta T$) may decrease or increase as a function of $\Delta T$ depending on the value of $\epsilon$. As the variations of $\mathcal{S}_1$ tend to be larger (in absolute value) than those of $\mathcal{S}_2$, the variations of the full noise are mainly due to the $\mathcal{S}_1$ contribution. This constitutes one of the central messages of this work on energy dependent transmission.

\begin{figure}[tbp]
	\centering
		\includegraphics[width=0.48\textwidth]{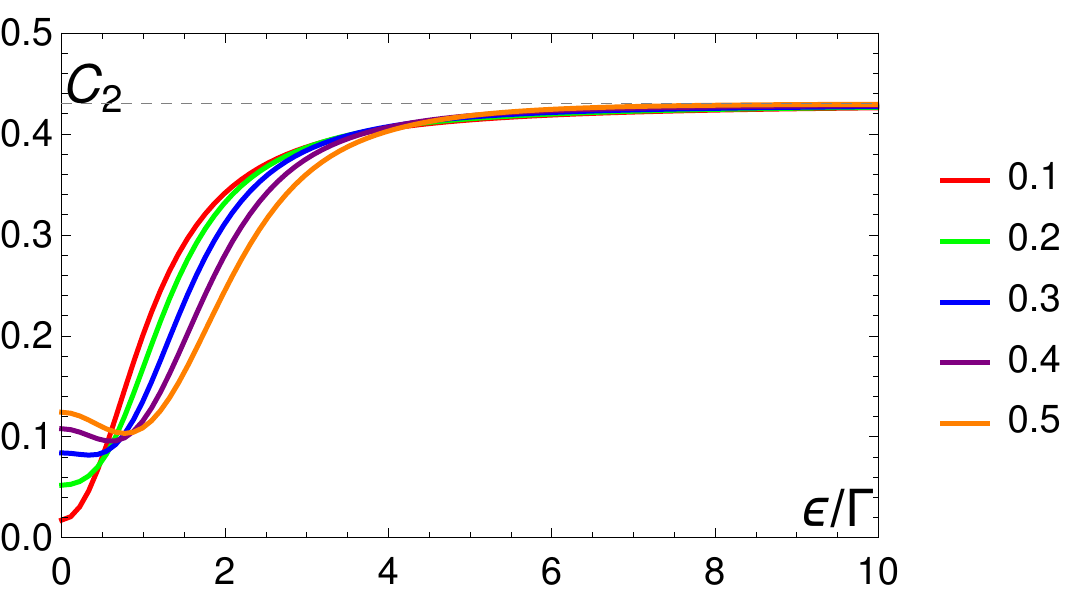}
		\includegraphics[width=0.48\textwidth]{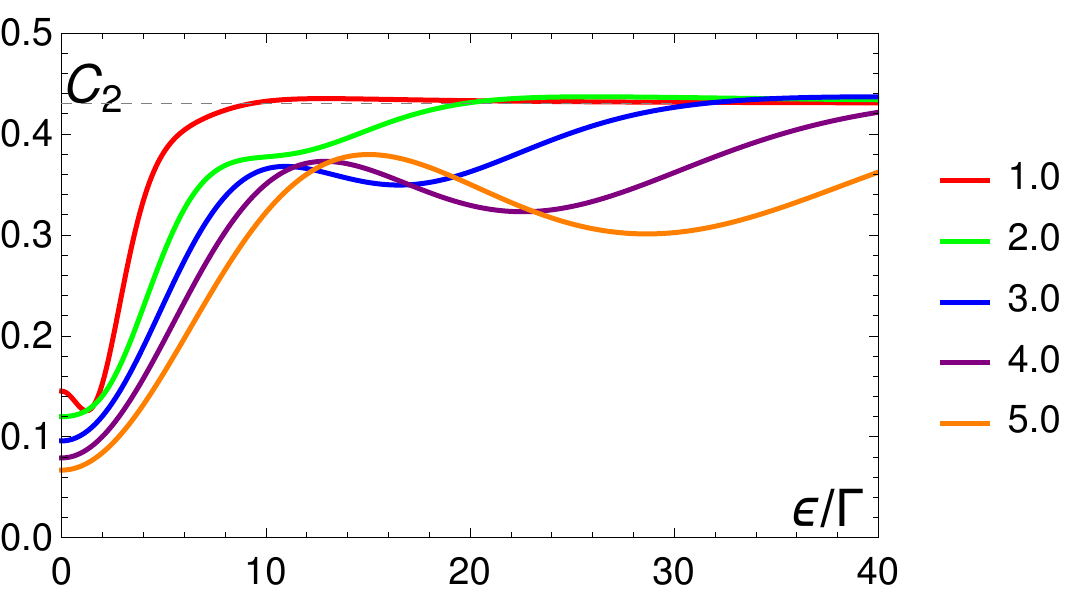}
		\caption{Coefficient $C_2$ of the $\mathcal{S}_2$ contribution as a function of $\epsilon/\Gamma$, for temperatures $\bar{T}/\Gamma=0.1,0.2,0.3,0.4,0.5$ (top) and $\bar{T}/\Gamma=1,2,3,4,5$ (bottom). The dashed line shows the value of Eq.~\eqref{c2constant} at $\tau\ll 1$ corresponding to a tunnel junction.}
		\label{fig:C2S2}
\end{figure} 

We now focus solely on the shot noise-like contribution $\mathcal{S}_2$, and on the coefficients of its expansion in powers of 
$\Delta T /(2 \bar {T})$:
\begin{equation}
\mathcal{S}_2 = \mathcal{S}_0 \left[C_2 \left( \frac{\Delta T}{2 \bar{T}}\right)^2 + C_4 \left( \frac{\Delta T}{2 \bar{T}}\right)^4 +  \cdots\right]~.
\label{eq:S2expansion}
\end{equation}

Fig.~\ref{fig:C2S2} (top) shows the value of $C_2$ as a function of $\epsilon/\Gamma$ for different values of $\bar{T}/\Gamma$ in the low temperature regime ($\bar{T} < \Gamma$). 
One can see that for the lowest displayed average temperatures, the coefficient $C_2$ has a minimum for $\epsilon=0$, where the transparency of the dot is 1. For such temperatures (lower two curves on the left), $C_2$ increases with $\epsilon$ in a monotonous way up to the value $\pi^2/9 - 2/3$ [see Eq.~\eqref{c2constant}] for a tunnel junction (see Sec.~\ref{comparable}; the tunnel limit is shown as the dashed line in Fig.~\ref{fig:C2S2}). This comes as no surprise.
Indeed, for $\epsilon>\Gamma$, the quantum dot embedded between the two normal leads plays the role of an adjustable QPC. This has been noticed in different contexts,\cite{jonckheere09,jacquet20} and the larger the $\epsilon$, the more this effective QPC becomes opaque and corresponds to a tunnel junction.
Increasing slightly the average temperature,  the minimum of $C_2$ shifts to $\epsilon\sim\Gamma$, and then $C_2$ increases until it saturates to the tunnel limit  ($\tau\ll 1$) of Eq.~\eqref{c2constant}. 

\begin{figure}[tbp]
	\centering
		\includegraphics[width=0.48\textwidth]{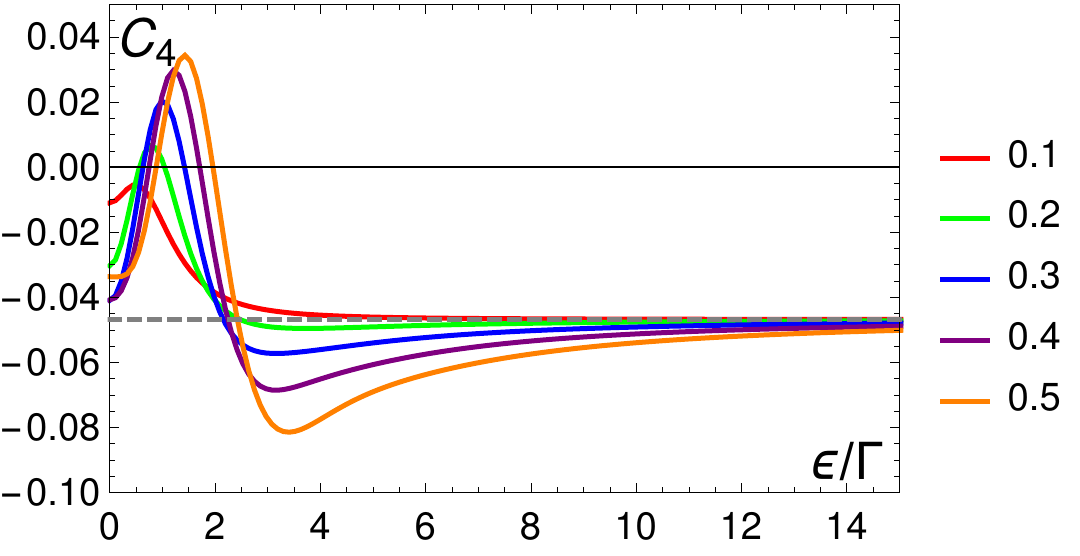}
		\includegraphics[width=0.48\textwidth]{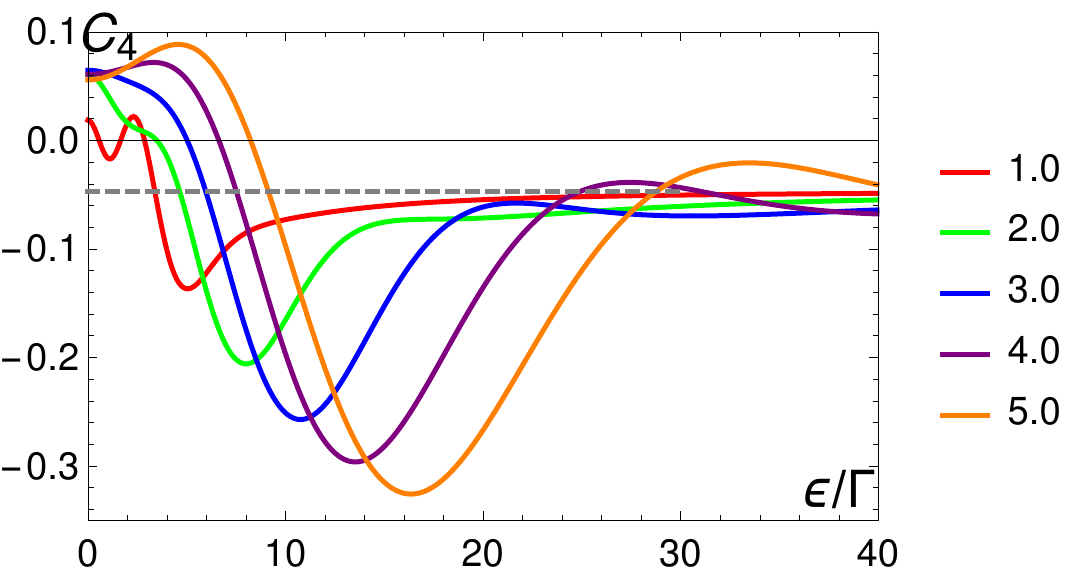}
		\caption{Coefficient $C_4$ of the $\mathcal{S}_2$ contribution as a function of $\epsilon/\Gamma$, for temperatures $\bar{T}/\Gamma=0.1,0.2,0.3,0.4,0.5$ (top) and  $\bar{T}/\Gamma=1,2,3,4,5$ (bottom). The dashed line shows the value of Eq.~\eqref{c4constant} at $\tau\ll 1$ corresponding to a tunnel junction.}
		\label{fig:C4S2}
\end{figure}

Similarly, Fig.~\ref{fig:C2S2} (bottom) shows the value of $C_2$ as a function of $\epsilon/\Gamma$ for different values of $\bar{T}/\Gamma$ in the high temperature regime ($\bar{T} > \Gamma$). For $\bar{T}/\Gamma=1$, $C_2$ first dips down to a global minimum before increasing to the saturation limit of a tunnel junction, as already observed in the low temperature regime. For $\bar{T}/\Gamma= 2, 3, 4, 5$, $C_2$ is minimal at $\epsilon=0$, then increases rapidly before slowly oscillating towards the saturation value of the tunnel limit for very large $\epsilon/\Gamma$.

While the measurement of the $C_4$ coefficient may be less relevant in experiments because of data accuracy, on the theory side it provides a control on the validity of the gradient expansion. Fig.~\ref{fig:C4S2} (top) shows the value of $C_4$ as a function of $\epsilon/\Gamma$ for different values of $\bar{T}/\Gamma$ in the low temperature regime ($\bar{T} < \Gamma$). In this regime, we observe that the overall amplitude of $C_4$ is small compared to that of $C_2$. For the lowest temperature, $C_4$ is always negative, it quickly reaches a global maximum before decreasing back towards the saturation value value of Eq.~\eqref{c4constant} at $\tau\ll 1$ corresponding to a tunnel junction. For higher temperatures in this plot, $C_4$ peaks up to a positive value, then dips down below the tunnel limit, before increasing back to reach the saturation value.  

Fig.~\ref{fig:C4S2} (bottom) shows the value of $C_4$ as a function of $\epsilon/\Gamma$ for different values of $\bar{T}/\Gamma$ in the high temperature regime ($\bar{T} > \Gamma$). This time we observe that the overall amplitude of $C_4$ is comparable to that of $C_2$, but as the gradient expansion is justified only for $\Delta T \ll \bar{T}$, we do not see a noticeable effect in the curves of $\mathcal{S}_2$ from Fig.~\ref{fig:Sdotfull}. At these higher temperatures, $C_4$ is positive for low values of $\epsilon/\Gamma$, but after some oscillations above/below the saturation level it reaches the tunnel limit, for all displayed temperatures. 

\subsection{Voltage and temperature biased case}
\label{voltage_bias}

We complete this section with the study of the combined effects of voltage bias and temperature gradient, persisting in the spirit of noise expansions in the temperature gradient.  We impose the voltage bias symmetrically on both leads as $V_1=-V_2=V/2$, which means that the breaking of electron-hole symmetry is still controlled by $\epsilon$. The calculations of the unbiased case can be straightforwardly  extended to include a voltage, by defining the excess noise as:
\begin{align}
\Delta \mathcal{S} (T_1,T_2,&V_1,V_2) = \mathcal{S}(T_1,T_2,V_1,V_2) \nonumber \\ &- 
  \frac{1}{2} \left[ \mathcal{S}(T_1,T_1,V_1,V_1) + \mathcal{S}(T_2,T_2,V_2,V_2) \right]~.
\end{align}
As in the absence of voltage noise, the excess noise contains only the non-equilibrium part of the noise,
and thus reduces to the $\mathcal{S}_2$ contribution.
In the regime of a small temperature bias, we can again perform a development in powers
of $\Delta T/(2 \bar{T})$, see Eq.~\eqref{eq:S2expansion}. Note that, when a bias voltage is applied
and when electron-hole symmetry is broken by using $\epsilon \neq 0$, the development also contains
odd powers of $\Delta T/(2 \bar{T})$. The coefficients are specified in Appendix \ref{appendix_comparable}.

\begin{figure}[tbp]
\centerline{\includegraphics[width=0.48\textwidth]{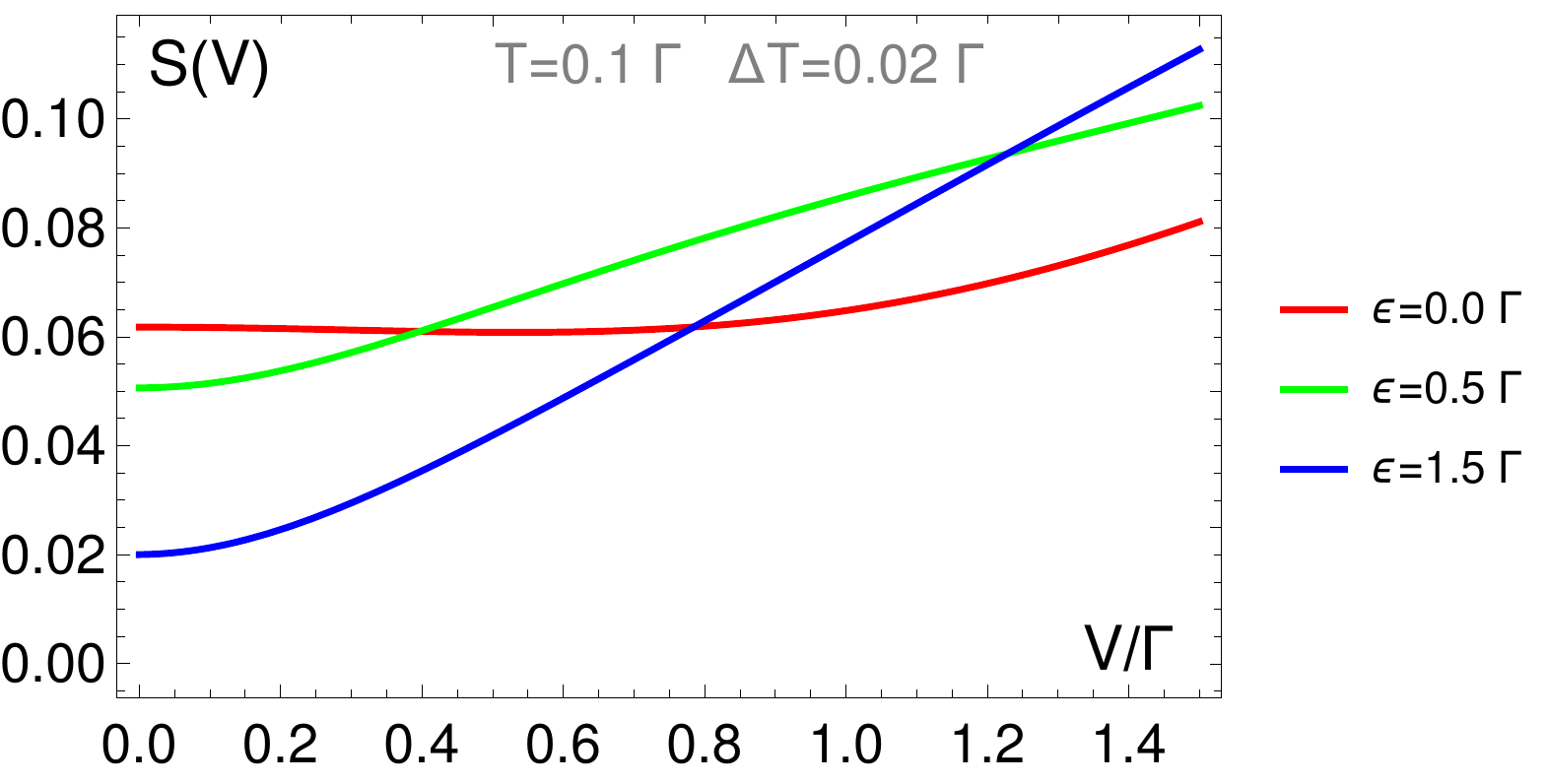}}
\caption{Full noise $\mathcal{S}=\mathcal{S}_1 + \mathcal{S}_2$ in units of $e^2 \Gamma$, as a function of the applied voltage $V/\Gamma$, for different
values of the resonant level position $\epsilon/\Gamma$, for a mean temperature $\bar{T}=0.1 \Gamma$ 
and a temperature gradient $\Delta T =0.02 \Gamma$.}
\label{fig:fullSofV}
\end{figure} 

Before showing the results for the coefficient $C_n$, it is instructive to look at the full noise
(containing the contributions $\mathcal{S}_1$ and $\mathcal{S}_2$) as a function of the bias voltage $V$, for a fixed temperature bias. This is shown for different values of $\epsilon$ in Fig.~\ref{fig:fullSofV} for the case of an average temperature $\bar{T} =0.1 \Gamma$
and a temperature bias $\Delta T = 0.02 \Gamma$. On the left of the figure, the value for $V=0$ corresponds
to the full DTN studied so far. 
The behavior when a finite voltage $V$ is applied is strongly dependent on the value
of $\epsilon$. Typically, the noise increases quickly with voltage, as it gets largely dominated by the shot noise contribution.
For $\epsilon=0$, it is nearly independent of $V$ up to $V \simeq \Gamma$, this is due to a competition between the two noise contributions as the transparency is close to 1 in this case. 

\begin{figure*}[tbp]
\includegraphics[width=0.96\textwidth]{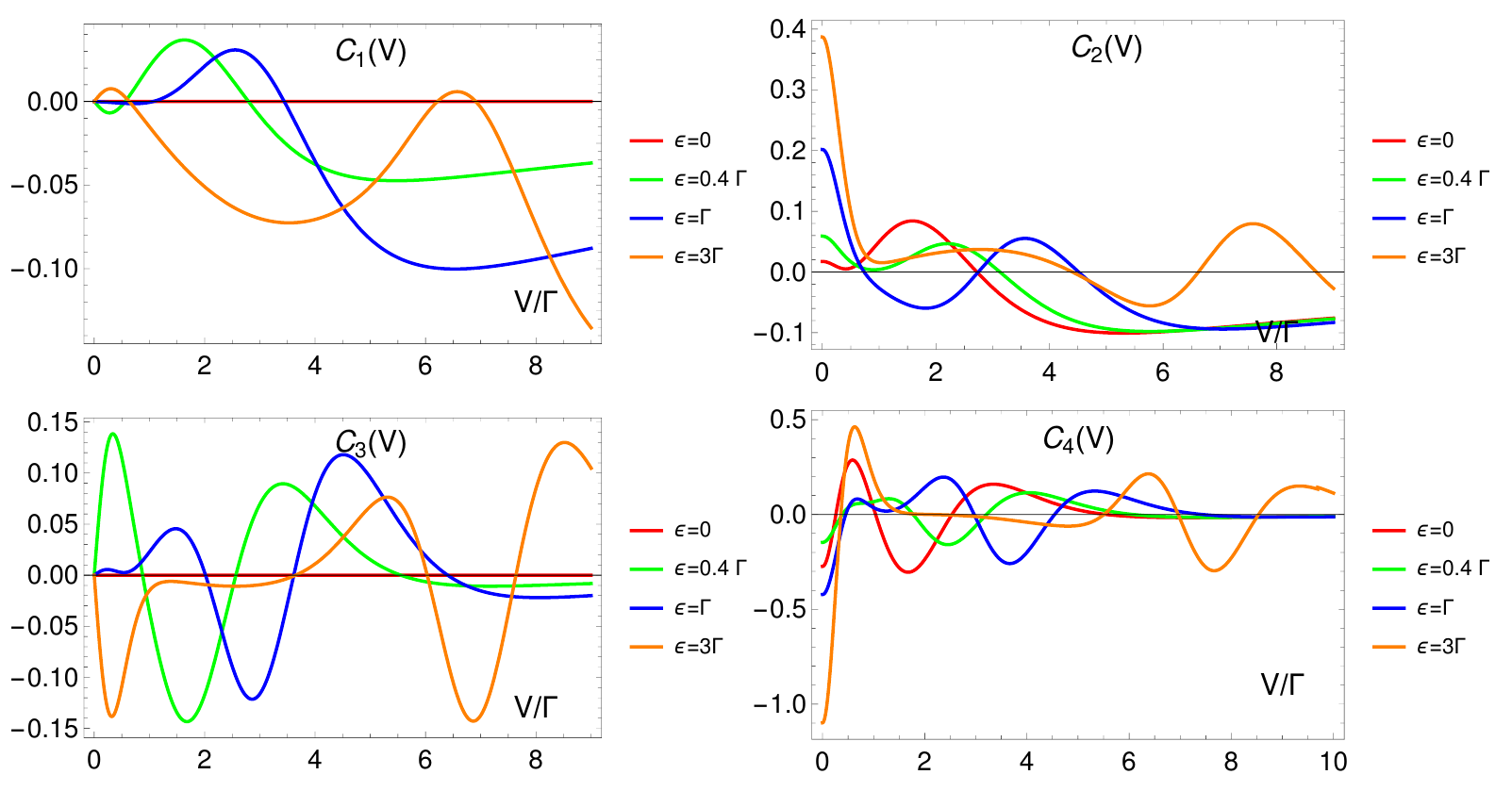}
\caption{Coefficients $C_1,C_2,C_3, C_4$ of the development of the excess noise in powers of
$\Delta T/(2 \bar{T})$ as a function of the bias voltage, for different values of the resonant level 
position $\epsilon$, for a temperature $T/\Gamma=0.1$.}
\label{fig:CV}
\end{figure*}

We now come to the coefficients of the $\Delta T$ dependence of the noise.
The behavior of the coefficients $C_1,C_2,C_3,C_4$ as a function of the voltage bias $V$
is shown in Fig.~\ref{fig:CV}. One can see on these four graphs that the bias voltage has a strong impact
on the coefficients, as it can induces sign changes and a complex behavior. The $C_1$ coefficient, which
is always zero in the absence of voltage bias, becomes non-zero when a voltage is applied and
the resonant level energy is non-zero, which means that the main dependence of the excess noise on the temperature bias becomes linear in this regime. Depending on the value of $V$, the coefficient $C_1$
can be positive or negative, but tends to be negative for larger values of $V$. Similarly, the coefficient $C_2$, which is always positive when no bias voltage is applied, can become negative when $V$ is large enough. The coefficients for higher orders, $C_3$ and $C_4$ also show complex oscillations as a function of the voltage.

\section{Breit-Wigner $\Delta T$ noise: cold reservoir case}
\label{breit_cold}

Still focusing on the case of a quantum dot described by a Breit-Wigner resonance as in Sec.~\ref{breit_comparable}, we consider here the opposite regime where the temperature bias is maximal, with $T_1=T$ and $T_2=0$. In this regime, we study the behavior of the excess noise $\mathcal{S}_2$ as the temperature $T$ of the hot reservoir is varied. We consider here two distinct regimes: the small (large) temperature  regime $T \ll \Gamma$ ($\Gamma \ll T$).

In the cold reservoir case, one of the two Fermi functions is assumed to be a step function.
From Eq.~\eqref{STN3}, the noise for a system under a bias $V$ (with $ \pm V/2$ applied on the right/left contact respectively) is then given by the integral :  
\begin{equation}
\mathcal{S}_2  = \frac{e^2}{\pi} \int_{-\infty}^{\infty} d\omega F(\omega) \left[ f\left(\omega-\frac{V}{2}\right) - \Theta\left(-\omega -\frac{V}{2}\right)\right]^2 \, ,
\label{hasegawa_start}
\end{equation}  
where
\begin{equation}
F(\omega)\equiv \tau(\omega) \left[1-\tau(\omega)\right]~,
\end{equation}
with the Breit-Wigner transmission $\tau(\omega)$, given in Eq.~\eqref{eq:BWformula}.
Note that in this section, no substantial complexity arises from further including a voltage bias in addition to the temperature gradient, so here we provide general expressions of the noise in the presence of both off-equilibrium conditions. We specify the unbiased case when relevant.

\subsection{Small temperature regime $\Gamma\gg T$}

We use the assumption that the typical width $\Gamma$ of $F(\omega)$
is much larger than $T$. One can then use a Taylor series of $F(\omega)$ in Eq.~\eqref{hasegawa_start} and exploit the properties of the Fermi function, which results into a development of  $\mathcal{S}_2$ in powers of $T/\Gamma$. In Appendix \ref{appendix_cold}, we derive analytical expressions of this expansion with the result:

\begin{widetext}
\begin{align}
\mathcal{S}_2=\frac{e^2}{\pi} & \left\{ (2 \ln 2 - 1) F\left(\frac{V}{2}  \right) \, T 
+ \sum_{m=1}^{\infty}  \left[(2 - 2^{1-2m})  \zeta(2m+1) - (2-2^{2-2m}) \zeta(2m)\right] F^{(2m)}\left(\frac{V}{2}\right)  \, T^{2m+1} \right. \nonumber\\ 
& \left. + \sum_{k=0}^{\infty}  c_k \left( \frac{V}{T} \right) F^{(k)} \left(\frac{V}{2}  \right)  \, T^{k+1}  \right\} ~,
\label{s_series}
\end{align}
\end{widetext}
where $F^{(n)}(\omega)= (\partial^n F/\partial \omega^n)$,  $\zeta(n)$ is the Riemann zeta function and the  $c_k \left( \frac{V}{T} \right)$ coefficients, which vanish at $V=0$  are specified in Appendix \ref{appendix_cold}..


We start with the unbiased case. In Ref.~\onlinecite{hasegawa21}, the first terms of the series of Eq.~\eqref{s_series} were derived for zero bias with the corresponding result:
\begin{align}
\mathcal{S}_2 = \frac{e^2}{\pi} &\left[ (2 \log{2}-1) F(0) T \right. \nonumber\\ 
&\left.+ \left(\frac{3}{2} \zeta(3) - \zeta(2)\right) F''(0) T^3 +...\right] ,
\label{eq:S2exp}
\end{align}
where the $\log{2}$ factor is reminiscent of the result of Ref.~\onlinecite{larocque20} which focuses on the total noise rather than the sole shot-noise-like contribution.
This equation shows that the generic behavior is $\mathcal{S}_2$ is
linear as a function of $T$, with a slope proportional to $\tau(0) \left[1-\tau(0) \right]$. This factor is typical for shot noise, and because the temperature is much smaller than the typical width $\Gamma$ of $\tau(\omega) \left[1-\tau(\omega)\right]$, it is only probed at $\omega=0$.  It is therefore only when $\tau(0) \left[1-\tau(0) \right]=0$ that the behavior of $\mathcal{S}_2$ is non-linear in $T$,
with a $T^3$ behavior. Importantly, this happens only at the dot resonance ($\epsilon=0$). 

For $\epsilon=0$, one thus has $\mathcal{S}_2(T,0)\propto T^3/\Gamma^2$, while off-resonance, for $\epsilon\neq 0$, $\mathcal{S}_2 (T,0) \propto \alpha T+\beta T^3$ is a superposition of a (positive) linear contribution and a cubic contribution, whose prefactor $\beta$ may change sign. This prefactor is related to the second derivative $F''(0)$, whose sign depends on the position of $\epsilon$ with respect to the inflection points of $F(\omega)$. Indeed, as $\omega$ and $\epsilon$ play a similar role in $F(\omega)$, we can   use the information of the Breit-Wigner transmission at $\epsilon= 0$ in order to understand the behavior of the excess noise when the dot position is shifted.  For $\epsilon=0$, $F(\omega)$ has two maxima at $\omega_{\pm}= \pm\Gamma$, and four inflection points at $\omega_{1 \pm}=\pm \sqrt{\frac{4-\sqrt{13}}{3}}  \Gamma$ and $\omega_{2 \pm}=\pm \sqrt{\frac{4 + \sqrt{13}}{3}}  \Gamma$.  It follows that when $\epsilon$ is below $\omega_{2-}$, above $\omega_{2+}$ or in the interval $\left[ \omega_{1 -} , \omega_{1 +} \right]$, $\beta$ is positive and changes sign otherwise.

\begin{figure}[tbp]
	\centering
		\includegraphics[width=0.48\textwidth]{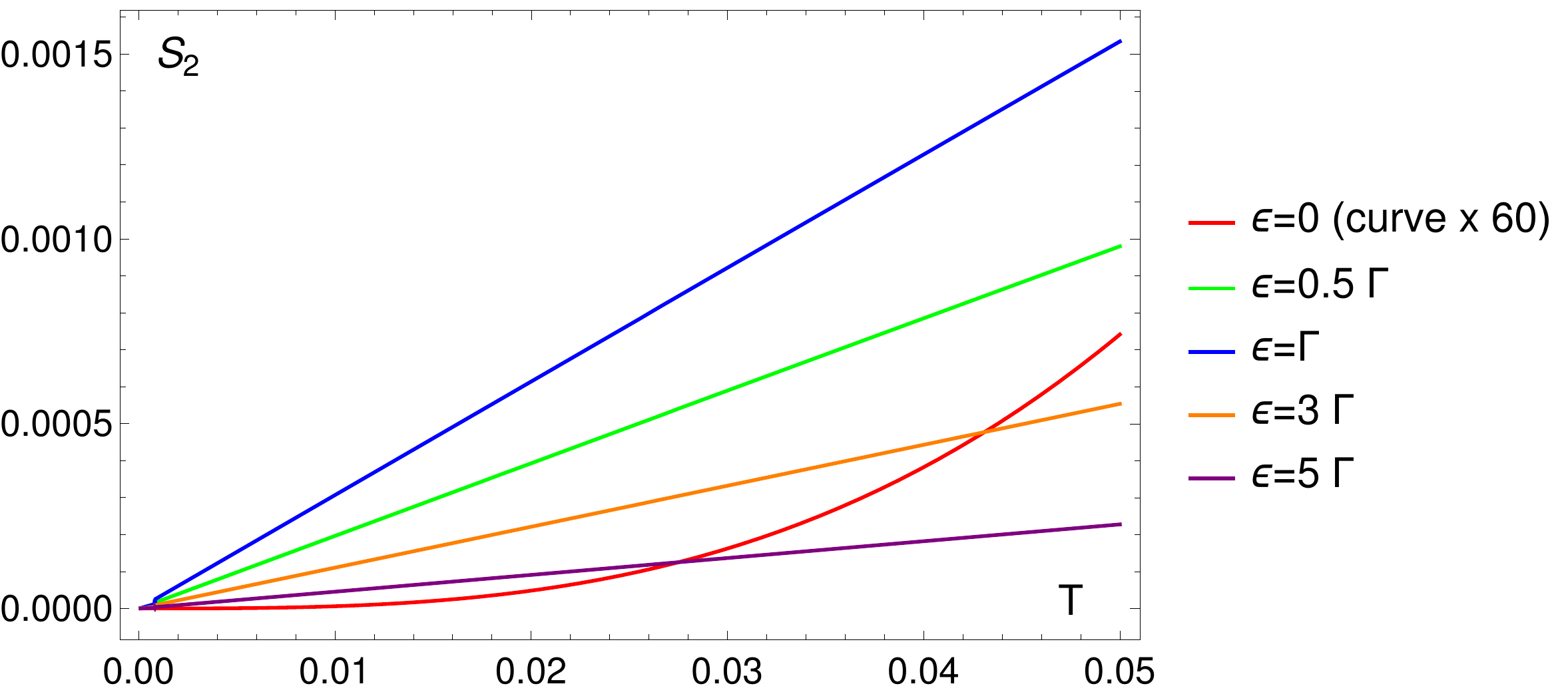}
		\caption{The $\mathcal{S}_2$ contribution to the noise, in units of $e^2 \Gamma$, as a function of temperature $T$ (in units of $\Gamma$)
		in the low temperature regime of the cold reservoir case ($T_1=T, T_2=0$),
		for different values of $\epsilon/\Gamma$ and in the absence of voltage bias, $V=0$. The curve for $\epsilon=0$ has been multiplied by 60 to be visible on the graph.}
		\label{fig:S2highT}
\end{figure}

The behavior of $\mathcal{S}_2(T)$ is illustrated in Fig.~\ref{fig:S2highT}, computed numerically from Eq.~\eqref{hasegawa_start} for several values
of $\epsilon/\Gamma$. For $\epsilon=0$, we observe the $T^3$ behavior, with a noise amplitude much smaller than the other cases 
obtained for $\epsilon \neq 0$ (note that the curve for $\epsilon=0$ has been multiplied by 60 
to be visible on the plot along with the other ones). The slope of the linear behavior is maximal for $\epsilon=\Gamma$,
and then decreases as $\epsilon$ gets larger and we reach the tunnel regime. This is consistent with the expression of Eq.~\eqref{eq:S2exp} where the slope of the linear in $T$ term is proportional to $F(0)$, with the typical $\log 2$ factor.\cite{larocque20}

One important conclusion of this section devoted to the cold reservoir case in the regime $T \ll \Gamma$ is that, contrarily to the situation where  both reservoirs have comparable temperatures, the excess noise contains only {\it odd} powers of the temperature difference $T$ (in the absence of a voltage bias). According to the small voltage expansion provided in Appendix \ref{appendix_cold}, even powers of $T$ might show up when $V\neq 0$, and these contributions are fully controllable.  Note however that given the starting assumption that $T\ll \Gamma$, the cubic contribution to $\mathcal{S}_2$ could only be detectable in experiments very close to or exactly on resonance. 

\subsection{Large temperature behavior}

We now consider the opposite limit  $T\gg \Gamma$. Our starting point is once again Eq.~\eqref{hasegawa_start}, where as a first step, we focus on the simpler case of a resonant level $\epsilon = 0$ and a vanishing voltage bias $V=0$. We thus obtain: 
\begin{equation}
\mathcal{S}_2= \frac{e^2}{2 \pi}\int_{0}^{+\infty} d\omega \frac{\Gamma^2 \omega^2}{\left( \Gamma^2 + \omega^2 \right)^2} \left[ \tanh \left( \frac{\omega}{2 T} \right) -1  \right]^2 ~.
\label{eq:resocold}
\end{equation}
In this situation, the Fermi function has slow variations compared to the sharp two-peak structure of $F(\omega)$ which shows maxima at $\omega_\pm= \pm \Gamma$. This allows us to approximate the above expression as follows
\begin{align}
\mathcal{S}_2 \simeq& \frac{e^2}{2 \pi}  \left[ \tanh \left( \frac{\Gamma}{2 T} \right) -1  \right]^2 \int_{0}^{+\infty} d\omega \frac{\Gamma^2 \omega^2}{\left( \Gamma^2 + \omega^2 \right)^2} \nonumber \\
\simeq & \frac{e^2}{2 \pi} \Gamma \frac{\pi}{4}~,
\label{saturation}
\end{align}
so that the shot-noise-like contribution saturates to a constant value at large enough temperature.

It is however interesting to try and go beyond this apparent saturation. Starting back from Eq.~\eqref{eq:resocold}, changing variables with $u = \omega/\Gamma$, then performing an expansion in $\Gamma/(2T)$ one is left with
\begin{equation}
\mathcal{S}_2 =  \frac{e^2}{2 \pi} \Gamma \left[ \frac{\pi}{4} +2 \frac{\Gamma}{2 T} \log \frac{\Gamma}{2 T} + O \left( \frac{\Gamma}{2 T} \right) \right]~.
\end{equation}
As it turns out, this first correction largely dominates, even when one deviates from resonance, or upon applying a voltage bias, provided that these satisfy $\left| \epsilon \right|, \left| V \right| \ll T$.

As a conclusion for the case $\Gamma\ll T$, one can obtain approximate, but analytic results for the excess noise $\mathcal{S}_2$.  Close to resonance, this noise saturates to the value: 
\begin{equation}
\mathcal{S}_2\approx \frac{\pi e^2\Gamma}{4 h}~,
\end{equation}  
which means that the quantum dot resonance defines some sort of a ``bias window'' of width of order $\Gamma$. This saturation is quite robust, even off-resonance,  as long as the temperature is much larger than both the dot energy level $\epsilon$ and voltage bias $V$, in addition to $\Gamma$.


\section{Conclusion}\label{conclusion}

We provided an historical perspective of its derivation in terms of equilibrium-like (${\mathcal S}_1$) and shot noise-like (${\mathcal S}_2$) contributions. This formula has been re-derived using two distinct approaches. First, the wave packet approach (for fermions) instigated by Landauer was reviewed. Next, we used the Keldysh formalism of non-equilibrium quantum field theory for a general tight binding model describing both QD and leads, which constitutes an extension of the derivation  of the current due to Caroli {\it et al.}.\cite{caroli71}. 

We summarized known results for a constant transmission coefficient when the reservoir temperatures are comparable, recalling that ${\mathcal S}_1={\mathcal S}_0$ does not depend on the temperature gradient. We stressed the additional relevance of the cold reservoir case, where the gradient expansion of DTN fails.

We moved on to generalize these results for an energy dependent transmission, first in the absence of bias. In this more general situation, the thermal-like contribution  ${\mathcal S}_1$ can be related to the zero bias differential conductance (which depends on either reservoir temperature) and its derivatives. The energy dependent transmission motivates to redefine the excess noise,\cite{hasegawa21} which turns out to correspond solely to ${\mathcal S}_2$. Indeed, the proper measurement of DTN should imply two equilibrium noise measurements at $T_1$ and $T_2$, followed by a non-equilibrium noise measurement in the presence of a temperature gradient.
Close to resonance, we pointed out the counter-intuitive result that for a given average temperature, the total noise ${\mathcal S}_1+{\mathcal S}_2$ {\it decreases} with an increasing temperature gradient, which is attributed to the dominance of the thermal contributions. 

We applied our results to the Breit-Wigner case of a single level QD. When the reservoir temperatures are comparable, the gradient expansion contains only even powers of the ratio $\Delta T/(2\bar{T})$. The coefficients $C_2$ and $C_4$ were expressed in terms of an energy integral implying higher derivatives of the distribution function [in the event that electron-hole symmetry is broken, the current bears a similar expansion, this time in terms of odd powers of  $\Delta T/(2\bar{T})$].
As a function of the dot level position, close to resonance, $C_2$ bears a minimum, and eventually increases in order to saturate to the expected tunnel limit ($C_4$ also saturates to its own tunnel limit far from resonance). $C_4$ has an amplitude which is small compared to that of $C_2$ for average temperatures which are small compared to the dot line width, but become comparable to $C_2$ for higher temperatures. At any rate, the gradient expansion seems justified when $\Delta T<\bar{T}$. The results are only slightly modified in the presence of a small bias voltage, giving us some confidence that a voltage generated by thermoelectric effect should not alter our conclusions.    

Another interesting limit concerns the situation where one of the reservoirs is set at zero temperature. We derived analytic expressions for the excess noise ${\mathcal S}_2$ in the form of infinite, odd power series in the ratio $T/\Gamma$ (which was assumed to be small). Odd powers such as $T^3$ dominate only when the dot is placed on resonance. Off resonance, we expect the excess noise to be linear in $T$ as in the case of a tunnel junction. On the other hand when $T\gg\Gamma$ (the opposite limit),  we showed that the excess noise saturates to a constant value which is independent of temperature, as if the width of the resonance played the role of a bias window.  

Along the course of this study, we noticed that Refs.~\onlinecite{mu19} and \onlinecite{eriksson21} addressed DTN and its generalization to include a voltage bias for a similar situation, with a Breit-Wigner transmission. While the results match with these works in the parameter range where there is overlap, our work mainly covers different regimes, and focuses on the evolution of different quantities than the ones explored in these studies.

The present results could be probed experimentally using an atomic break junction with a ``molecule'' embedded between the two leads, which plays the role of a QD. Care should be taken to use a molecule which couplings to the leads are sufficiently weak as to define a QD with quantized energy levels (thus ruling out the use of an hydrogen molecule which has the opposite property). However in such a setup, it is typically difficult to approach the molecule with a side gate, which is necessary to probe the on/off resonance condition (an overall back gate would be necessary to tune the dot level). Alternatively, one could work with a two-dimensional electron gas defined in a semi-conductor heterostructure, with two QPC defining the dot as in Fig \ref{fig00}, and side gates to tailor the properties of the QD level.

Possible extensions of this work include: 

a) Accounting for electronic correlations in the QD: if this dot has a small capacitance, Coulomb blockade effects, or even Kondo physics could be at play. Scattering theory cannot access such phenomena straightforwardly and one could possibly resort to a perturbative treatment of the Coulomb interaction on the QD using the Keldysh formalism, albeit in a current conserving manner. This was achieved previously by some of the authors in the context of hybrid superconducting devices.\cite{rech12} For the Kondo regime, there exists a recent study.\cite{hasegawa21}

b) The extension to temperature and voltage biased multichannel, multiterminal devices,\cite{hajiloo20} which would require a substantial dose of linear algebra, as in Refs. \onlinecite{buttiker90,buttiker92} for the purely voltage biased case. In full generality, many of the previous studies on noise in mesoscopic/nanoscopic systems could be repeated for an unbiased system subject to a temperature gradient.  For instance, as two particle interference effects were predicted to appear in the noise of voltage biased devices,\cite{samuelsson04,splettstoesser09} with connections to non-local entanglement. It would be interesting to probe whether such effects also occur when a temperature gradient is imposed, as usually the temperature  typically contributes to decoherence effects.  
 
c) The study of  DTN at finite frequencies: in voltage biased ballistic mesoscopic systems such as a QPC, finite frequency excess noise has a singular derivative when the frequency matches the voltage. It would be pertinent to see if the noise spectrum bears peculiar features associated with the presence of a temperature gradient. 

\acknowledgments
This work has been carried out under COVID-19 circumstances with teleworking conditions. One of the authors (TM) dedicates this work to the memory of Andr\'e Martin, physicist from CERN. The project leading to this publication has received funding from Excellence Initiative of Aix-Marseille University - A*MIDEX, a French ``Investissements d'Avenir'' program through the IPhU (AMX-19-IET-008) and AMUtech (AMX-19-IET-01X) institutes.

\appendix

\section{Wavepacket approach to the noise formula}
\label{sec:wavepacket_appendix}
\subsection{Two fermion collision in the wave packet approach}

Here, we argue that the Pauli principle prohibits the double occupancy of outgoing states, and we compute the amplitude of other outcomes where the two electrons are either transmitted or reflected in opposite reservoirs. Special attention is taken for the case where two electrons are incident - from the right and from the left - with occupancy $f_1 f_2$. 

Naively speaking, if electrons were classical objects, one would expect the following $4$ possibilities:

a) both electrons end up on the right, with an ``apparent'' probability $P_a=\tau f_1(1-\tau)f_2$.

b) both electrons end up on the left, with an ``apparent'' probability $P_b=(1-\tau) f_1 \tau f_2$.

c) both electrons are transmitted with probability $P_c=\tau^2f_1f_2$.

d) both electrons are reflected with probability $P_d =(1- \tau)^2f_1f_2$.

In Refs. \onlinecite{landauer89,landauer91,martin92} it was argued correctly that processes a) and b) are forbidden by the Pauli principle because the two outgoing electrons end up in the same state. We argue here that indeed the amplitudes  $A_a$ and $A_b$ (rather than probabilities) associated with a) and b) vanish because the wave function associated with the two electrons remains antisymmetric, throughout the evolution, which yields $P_a=P_b=0$. The fact that $P_c + P_d\neq f_1f_2$ is worrisome, as the sum of the probabilities for these processes should satisfy this relation.

However, the reasoning behind events c) and d) is incorrect: one cannot dissociate these two events, because the incoming and outgoing states are the same for c) and d) and the two incoming electrons are identical particles.
The two corresponding amplitudes $A_c$ and $A_d$ are shown here to add up to one, apart from a complex phase factor. 

The scattering matrix associated with the evolution of electrons traveling through the scattering region reads:
\begin{equation}
    s=\begin{pmatrix}
    r&t\\
    t&r'\\
    \end{pmatrix}
    = \begin{pmatrix}
    \sqrt{1-\tau}\text{e}^{\text{i}\phi'}& \sqrt{\tau}\text{e}^{\text{i}\phi}\\
    \sqrt{\tau}\text{e}^{\text{i}\phi} & -\sqrt{1-\tau}\text{e}^{-\text{i}\phi'+2\text{i}\phi}
    \end{pmatrix}~.
\end{equation}

\begin{figure}[tbp]
	\centering
		\includegraphics[width=0.48\textwidth]{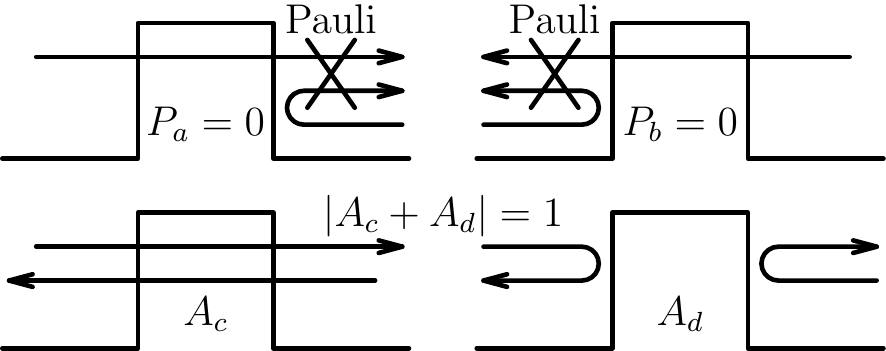}
		\caption{Situation where two fermions are incident from both leads, with occupancy probability $f_1f_2$. Top: processes $a$ and $b$ which are forbidden by the Pauli principle, and whose amplitude vanishes. Bottom: indistinguishable processes $c$ and $d$ which lead to zero charge transfer and whose total amplitude has an absolute value equal to $1$.}
		\label{fig:collision}
\end{figure}

For fermions, the wave function of the two incoming electrons (from opposite sides of the sample) is antisymmetric:

\begin{equation}
\Psi_i(x_1,x_2) = \frac{1}{\sqrt{2}}(1-\hat{P}_{21}) \varphi_{i_1}(x_1)  \varphi_{i_2}(x_2)~,
\end{equation}
where $\hat{P}_{21}$ is the permutation operator, $\varphi_{i_1}(x_1)$ and $\varphi_{i_2}(x_2)$  are the single particle wave functions of the incoming states at the two outputs.

Similarly, the outgoing state is also anti-symmetrized
\begin{equation}
\Psi_o(x_1,x_2)=\frac{1}{\sqrt{2}}(1-\hat{P}_{21})\varphi_{o_j}(x_1)  \varphi_{o_k}(x_2)~,
\end{equation}
with $j,k=1,2$. The anti-symmetrized two-particle state has no weight if the two electrons end up in the same output state ($j=k$), a direct reflection of the Pauli principle. One concludes that the final state has $j=1$, $k=2$ or $j=2$, $k=1$. 

In order to compute the amplitude associated with 2 electrons being both transmitted or both reflected, we need to consider the overlap between the final state and the (evolved) initial state:     
\begin{align}
A_c+A_d \equiv& \langle \Psi_o  | U  | \Psi_i \rangle\nonumber\\
 =& \int dx_1 dx_2 \varphi^*_{o_1}(x_1) \varphi^*_{o_ 2}(x_2)   (1-\hat{P}_{21})  \nonumber \\
 & \qquad \times U  \varphi_{i_1} (x_1) \varphi_{i_2}(x_2)~,
\end{align}
where we used the standard assumption that $\hat{P}_{21}$ commutes with the two particle evolution operator $U$ (a tensor product of the scattering matrices), which yields:
\begin{align}
A_c+A_d =& \int dx_1 dx_2 \varphi^*_{o_1}(x_1) \varphi^*_{o_ 2}(x_2)   (1-\hat{P}_{21})\nonumber\\ 
& \quad \times \left[ tt' \varphi_{o_2}(x_1) \varphi_{o_1}(x_2) + rr' \varphi_{o_1}(x_1) \varphi_{o_2}(x_2) \right] \nonumber\\
=& -tt'+rr' \nonumber \\
=& e^{i\pi + 2i\phi}~,
\end{align}
where we used the orthogonality relation between opposite outgoing states.

The conclusion is that the doubly occupied state which corresponds to zero net current $g=0$ has the probability: 
\begin{equation}
P_{c+d}= | A_c+A_d |^2 f_1f_2=f_1f_2 ~.
\end{equation}

The variance of Eq.~\eqref{noisevariance} has only contributions from $g=\pm 1$:
\begin{equation}
\left\langle g^2 \right\rangle - \left\langle g \right\rangle^2 = \tau f_1(1-f_2) + \tau f_2(1-f_1) - \tau^2 (f_1-f_2)^2~,
\end{equation}
and the STN formula of Eq.~\eqref{eq:STNformula} is recovered.

\subsection{Two boson collision}

It is interesting to compare the previous result for two bosons incoming  from  different reservoirs, where the incoming  and outgoing two-particle state is symmetric.  
\begin{equation}
\Psi_i(x_1,x_2)=\frac{1}{\sqrt{2}}(1+\hat{P}_{21})\varphi_{i_1}(x_1)  \varphi_{i_2}(x_2) ~,
\end{equation}
and similarly for the two particle outgoing state, which has now three outcomes:

a) both bosons end up in different outputs: $\Psi_o(x_1,x_2)=\frac{1+\hat{P}_{21}}{\sqrt{2}}\varphi_{o_1}(x_1)  \varphi_{o_2}(x_2)$  

b) both bosons end up in the same state on the right: $\Psi_o(x_1,x_2)=\varphi_{o_2}(x_1)  \varphi_{o_2}(x_2)$

c) both bosons end up on the left: $\Psi_o(x_1,x_2)=\varphi_{o_1}(x_1)  \varphi_{o_1}(x_1)$ 

A calculation similar to that of fermions leads to the result:
\begin{align}
A_a =& \int dx_1 dx_2 \varphi^*_{o_1}(x_1) \varphi^*_{o_ 2}(x_2)   (1+\hat{P}_{21}) \nonumber\\
& \quad \times \left[ tt' \varphi_{o_2}(x_1) \varphi_{o_1}(x_2) + rr' \varphi_{o_1}(x_1) \varphi_{o_2}(x_2) \right] \nonumber\\
=& tt'+rr' \nonumber \\
=& (2\tau-1)e^{i\phi}~.
\end{align}
The associated probability is: 
\begin{equation}
P_{a}= | A_a|^2 f_1f_2= (4\tau^2+4\tau +1) f_1f_2~.
\end{equation}

When the two bosons end up on the right:
\begin{align}
A_b=& \int dx_1 dx_2 \varphi^*_{o_2}(x_1) \varphi^*_{o_ 2}(x_2)   \frac{1+\hat{P}_{21}}{\sqrt{2}}tr'   \varphi_{o_2}(x_1) \varphi_{o_2}(x_2) \nonumber\\
=& \sqrt{2} tr'~.
\end{align}
When two bosons end up on the left :
\begin{equation}
A_c= \sqrt{2} t'r ~,
\end{equation}
and the associated probabilities are:
\begin{equation}
P_b = P_c =2\tau(1-\tau)f_1f_2~.
\end{equation}
The presence of the extra factor $2$ reflects the bunching character of bosons. Note that $\vert A_a\vert^2+\vert A_b\vert^2+\vert A_c\vert^2=1$ as expected.

\section{Microscopic derivation} \label{app:micro}

Here we provide some additional details concerning the derivations of Sec.~\ref{sec:microscopic}.

Working out the product of matrices of Eq.~\eqref{eq:Gpmmp} explicitly in the site basis, one has
\begin{align}
\label{eq:GaaPM}
G_{a a}^{\pm \mp} =& \lambda_a^2 G_{a a}^R  G_{a a}^A g_\alpha^{\pm \mp} + \lambda_b^2 G_{a b}^R  G_{b a}^A  g_\beta^{\pm \mp} , \\
\label{eq:GalphaalphaPM}
G_{\alpha \alpha}^{\pm \mp} =& \left( 1 + \lambda_a G_{\alpha a}^R \right)  \left( 1 + \lambda_a G_{a \alpha}^A \right) g_\alpha^{\pm \mp} + \lambda_b^2 G_{\alpha b}^R  G_{b \alpha}^A  g_\beta^{\pm \mp} , \\ 
G_{\alpha a}^{\pm \mp} =& \left( 1 + \lambda_a G_{\alpha a}^R \right)  \lambda_a G_{a a}^A  g_\alpha^{\pm \mp} + \lambda_b^2 G_{\alpha b}^R  G_{b a}^A  g_\beta^{\pm \mp} , \\ 
G_{a \alpha}^{\pm \mp} =&  \lambda_a G_{a a}^R  \left( 1 + \lambda_a G_{a \alpha}^A \right) g_\alpha^{\pm \mp} + \lambda_b^2 G_{a b}^R  G_{b \alpha}^A  g_\beta^{\pm \mp} ,
\end{align}
where we used that all contributions involving $g_{aa}^{\pm \mp}, g_{ab}^{\pm \mp}, g_{ba}^{\pm \mp}$ and $g_{bb}^{\pm \mp}$ vanish (see Ref.~\onlinecite{rechprivate}). Using the first two expressions, the last two can be further rewritten as
\begin{align}
\label{eq:newGalphaaPM}
G_{\alpha a}^{\pm \mp} =& \lambda_a \left( g_{\alpha}^{\pm \mp} G_{aa}^A + g_{\alpha}^R G_{aa}^{\pm \mp} \right) ~, \\
G_{a \alpha}^{\pm \mp} =& \lambda_a \left( g_{\alpha}^{A} G_{aa}^{\pm \mp} + g_{\alpha}^{\pm \mp} G_{aa}^R \right) ~.
\label{eq:newGaalphaPM}
\end{align}

Combining Eq.~\eqref{eq:newGalphaaPM} and Eq.~\eqref{eq:newGaalphaPM}, one can write
\begin{align}
G_{\alpha a}^{+-}  - G_{a \alpha}^{+-}  =  \lambda_a & \left[ 
g_{\alpha}^{+-} \left( G_{aa}^A - G_{aa}^R \right) \right. \nonumber \\
& \left. + \left( g_{\alpha}^R - g_{\alpha}^{A} \right) G_{aa}^{+-} 
\right] ~.
\end{align}
Keeping in mind that $G^A - G^R = G^{+-} - G^{-+}$ and $g^R - g^A = g^{-+} - g^{+-}$, this can be further reduced to
\begin{align}
G_{\alpha a}^{+-}  - G_{a \alpha}^{+-} 
=&  \lambda_a \left[ 
 g_{\alpha}^{-+}  G_{aa}^{+-} - g_{\alpha}^{+-} G_{aa}^{-+}
\right] \nonumber \\
=&  \lambda_a  \lambda_b^2 G_{a b}^R  G_{b a}^A \left[ 
   g_{\alpha}^{-+}   g_\beta^{+-}
  - g_\beta^{-+}  g_{\alpha}^{+-}
\right] ~,
\label{eq:Galphaa-Gaalpha}
\end{align}
where we used the expression for $G_{aa}^{\pm \mp}$ given in Eq.~\eqref{eq:GaaPM}. Substituting then the expression for the bare lead Green's functions leads back to the formula for the mean current quoted in the main text, Eq.~\ref{eq:microcurrentformula}.

Using Eqs.~\eqref{eq:newGalphaaPM}-\eqref{eq:newGaalphaPM} for $G_{\alpha a}^{\pm \mp}$ and $G_{a \alpha}^{\pm \mp}$ along with Eq.~\eqref{eq:GaaPM} for $G_{a a}^{\pm \mp}$, one has
\begin{align}
\label{eq:newGalphaaPM2}
G_{\alpha a}^{\pm \mp} =& 
\lambda_a G_{aa}^A \left( 1 + \lambda_a^2 g_{\alpha}^R G_{a a}^R \right) g_\alpha^{\pm \mp} 
+ \lambda_a \lambda_b^2 G_{a b}^R  G_{b a}^A g_{\alpha}^R g_\beta^{\pm \mp}  ~, \\ 
G_{a \alpha}^{\pm \mp} =& 
\lambda_a G_{aa}^R \left( 1 + \lambda_a^2 g_{\alpha}^{A} G_{a a}^A \right) g_\alpha^{\pm \mp} 
+ \lambda_a  \lambda_b^2 G_{a b}^R  G_{b a}^A  g_{\alpha}^{A}  g_\beta^{\pm \mp} ~.
\label{eq:newGaalphaPM2}
\end{align}

Combining these two with Eq.~\eqref{eq:GaaPM} for $G_{a a}^{\pm \mp}$ and Eq.~\eqref{eq:GalphaalphaPM} for $G_{\alpha \alpha}^{\pm \mp}$, one can rewrite the integrand of Eq.~\eqref{eq:Saadef} for the zero frequency noise as
\begin{align}
{\cal I}_N =& G_{\alpha a}^{-+} G_{\alpha a}^{+-}  -  G_{\alpha \alpha}^{- +}  G_{a a}^{+-}  -  G_{a a}^{-+} G_{\alpha \alpha}^{+ -} +  G_{a \alpha}^{- +}  G_{a \alpha}^{+ -} \nonumber \\
=& \lambda_a^2 \left[ G_{a a}^A - G_{a a}^R +\lambda_a^2 \left( g_\alpha^R - g_\alpha^A \right) G_{a a}^A G_{a a}^R \right]^2 g_\alpha^{-+} g_\alpha^{+-} 
\nonumber \\
& + \lambda_a^2 \lambda_b^4 \left( G_{ab}^R G_{ba}^A \right)^2 \left( g_\alpha^A - g_\alpha^R \right)^2 g_\beta^{-+} g_\beta^{+-}  \nonumber \\
 & + \left[ -1 +\lambda_a^2 \left( g_\alpha^R - g_\alpha^A \right) G_{a a}^A  \right] 
\left[ 1 +\lambda_a^2 \left( g_\alpha^R - g_\alpha^A \right) G_{a a}^R  \right] \nonumber \\
& \quad \times  \lambda_b^2 G_{ab}^R G_{ba}^A \left( g_\alpha^{-+} g_\beta^{+-} + g_\beta^{-+} g_\alpha^{+-} \right) ~.
\end{align}
At this stage, it is useful again to notice that  $G^A - G^R = G^{+-} - G^{-+}$ and $g^{+-} - g^{-+} = g^A - g^R$, so that one has
\begin{align}
G_{a a}^A - G_{a a}^R &= G_{a a}^{+-} - G_{a a}^{-+} \nonumber \\
&= \lambda_a^2 G_{a a}^R  G_{a a}^A \left( g_\alpha^A - g_\alpha^R \right) \nonumber \\
& \qquad + \lambda_b^2 G_{a b}^R  G_{b a}^A  \left( g_\beta^A - g_\beta^R \right) ~,
\end{align}
which allows to write
\begin{align}
G_{a a}^A - G_{a a}^R  +\lambda_a^2 \left( g_\alpha^R - g_\alpha^A \right) & G_{a a}^A G_{a a}^R \nonumber \\
&=  \lambda_b^2 G_{ab}^R G_{ba}^A \left( g_\beta^A - g_\beta^R \right)   ~,
\end{align}
as well as
\begin{align}
&\left[ -1 +\lambda_a^2 \left( g_\alpha^R - g_\alpha^A \right) G_{a a}^A  \right] 
\left[ 1 +\lambda_a^2 \left( g_\alpha^R - g_\alpha^A \right) G_{a a}^R  \right] \nonumber \\
&\qquad = -1 + \lambda_a^2 \lambda_b^2 G_{ab}^R G_{ba}^A \left( g_\beta^A - g_\beta^R \right) \left( g_\alpha^R - g_\alpha^A \right)  ~,
\end{align}
ultimately leading to 
\begin{align}
{\cal I}_N =& \lambda_a^2 \lambda_b^4 \left( G_{ab}^R G_{ba}^A \right)^2 \nonumber \\
& \quad \times \left[ \left( g_\beta^A - g_\beta^R \right)^2 g_\alpha^{-+} g_\alpha^{+-}  +  \left( g_\alpha^A - g_\alpha^R \right)^2 g_\beta^{-+} g_\beta^{+-} \right] \nonumber \\
& - \left[ 1 + \lambda_a^2 \lambda_b^2 G_{ab}^R G_{ba}^A \left( g_\beta^A - g_\beta^R \right) \left( g_\alpha^A - g_\alpha^R \right)  \right] \nonumber \\
& \quad \times \lambda_b^2 G_{ab}^R G_{ba}^A \left( g_\alpha^{-+} g_\beta^{+-} + g_\beta^{-+} g_\alpha^{+-} \right) ~,
\end{align}
which is the expression quoted in the main text, Eq.~\eqref{eq:micronoise}, after substituting the bare lead Green's functions.

\section{$\Delta T$ noise in the presence of a voltage}
\label{appendix_voltage}

In this appendix, we provide formulas for the voltage dependence of DTN in the presence of a voltage bias which respects electron-hole symmetry, both when the reservoir temperatures are comparable and when one reservoir is set to zero temperature. 

\subsection{Comparable temperatures}
\label{appendix_comparable}

\begin{widetext}
\begin{align}
C_0 &=\frac{e^2}{{\pi\mathcal S}_0} \int d\omega \;    \tau(\omega)\left[1-\tau(\omega)\right]
\left( \sum_{\sigma=\pm} \sigma{\bar f}_\sigma \right)^2 ~, 
\\
C_1 &=  \frac{\bar{T}e^2}{\pi{\mathcal S}_0}   \int d\omega \; \tau(\omega) \left[1-\tau(\omega)\right]
\left( 2 \sum_{\sigma, \sigma' =\pm} \sigma{\bar f}_\sigma \times \partial_{\bar{T}} {\bar f}_{\sigma'}  \right) ~,
\\
C_2 &=  \frac{\bar{T}^2e^2}{\pi{\mathcal S}_0}   \int d\omega \; \tau(\omega) \left[1-\tau(\omega)\right]
\left[ \left(\sum_{\sigma=\pm} \partial_{\bar{T}}{\bar f}_\sigma\right)^2    +      \sum_{\sigma, \sigma' =\pm} \sigma{\bar f}_\sigma \times  \sigma'\partial^2_{\bar{T}}{\bar f}_{\sigma'} \right] ~,
\\
C_3&=  \frac{\bar{T}^3e^2}{\pi{\mathcal S}_0}   \int d\omega \; \tau(\omega) \left[1-\tau(\omega)\right] \left[  \sum_{\sigma, \sigma' =\pm} \left( \partial_{\bar{T}} {\bar f}_\sigma \times \sigma'\partial^2_{\bar{T}} {\bar f}_{\sigma'} + \frac{1}{3} \sigma{\bar f}_\sigma \times \partial^3_{\bar{T}}{\bar f}_{\sigma'} \right) \right] ~,
\label{c3energy}\\
C_4&=  \frac{\bar{T}^4e^2}{\pi{\mathcal S}_0}  \int d\omega \; \tau(\omega) \left[1-\tau(\omega)\right]
\left[ 
\frac{1}{4}\left( \sum_{\sigma=\pm} \sigma \partial^2_{\bar{T}} {\bar f}_\sigma \right)^2  +  \sum_{\sigma, \sigma' =\pm} \left( \frac{1}{3} \partial_{\bar{T}}{\bar f}_\sigma\times \partial^3_{\bar{T}}{\bar f}_{\sigma'} + \frac{1}{12} \sigma{\bar f}_\sigma \times \sigma'\partial^4_{\bar{T}} {\bar f}_{\sigma'} \right) \right] ~,
\end{align}
\end{widetext}
where ${\bar f}_\sigma$ are the lead distributions evaluated at $\bar{T}$ ($f_{+}$ for lead 1 at voltage
$V/2$, $f_{-}$ for lead 2 at voltage $-V/2$). Note that the presence of the voltage bias implies that there is a finite shot noise for $\Delta T =0$, which leads to a coefficient $C_0\neq 1$. The integrand in these expressions can be cast into energy derivatives using the relation $\partial_T{\bar f}_\sigma =-\frac{\omega-\sigma V/2}{T}\partial_\omega{\bar f}_\sigma$. We obtain similar expressions for the current expansion coefficients:
\begin{equation}
D_n =\frac{e}{2\pi}\int d\omega \tau(\omega) \frac{1}{n!} \sum_{\sigma=\pm}  \left (-\frac{\omega-\sigma V/2}{T}\partial_\omega\right)^n \sigma{\bar f}_\sigma ~.
\label{dn}
\end{equation}


\subsection{Cold reservoir}
\label{appendix_cold}

We define $I =  \frac{\pi}{e^2} \mathcal{S}_2$. We wish to evaluate this integral using a Taylor expansion of $F(\omega)$ in frequency.

We first remark that 
\begin{align}
\left[ f\left( \omega - \frac{V}{2}\right) - \Theta\left(-\omega -\frac{V}{2}\right) \right]^2   
 =&  T \frac{df}{d\omega} \left(\omega-\frac{V}{2}\right)\nonumber\\
 & + g\left(\omega-\frac{V}{2}\right) \, ,
\end{align}
where $g(\omega) = f(\omega) -2 f(\omega)\Theta(-\omega-V) +\Theta(-\omega -V)$.

We decompose $I$ in two terms $I = I_1+ I_2$, and we evaluate these two contributions separately. 
\begin{equation}
I_1 =T \int_{-\infty}^{\infty} d \omega F\left(\omega +\frac{V}{2}  \right) \frac{df}{d\omega} \left(\omega\right) \, ,
\end{equation}
where we may substitute $F\left(\omega +\frac{V}{2}  \right)$ by its Taylor expansion 
\begin{equation}
F\left(\omega +\frac{V}{2}  \right) = \sum_{k=0}^{\infty} F^{(k)} \left(\frac{V}{2} \right)\frac{\omega^k}{k!} \,,
\end{equation}
so that
\begin{equation}
I_1 = -T F\left(\frac{V}{2}  \right) + T \sum_{k=1}^{\infty} \frac{1}{k!} F^{(k)} \left(\frac{V}{2}  \right) \int_{-\infty}^{\infty} d \omega \omega^k \frac{df}{d\omega} (\omega) \, .
\end{equation}

We  thus have to compute for $k>0$: 
\begin{equation}
T \int_{-\infty}^{\infty} d \omega \, \,  \omega^k \frac{df}{d\omega} \left(\omega\right) 
= \frac{-T^{k+1}}{4} \int_{-\infty}^{\infty} du  \frac{u^k}{\mathrm{cosh}^2(u/2)} \,.
\end{equation}

When $k$ is odd the integral vanishes, whereas for even $k = 2m$ it reads (see Ref.~\onlinecite{gradshteyn14}, 3.527.3)
\begin{equation}
 \int_{0}^{\infty} du  \frac{u^{2m}}{\mathrm{cosh}^2(u/2)} =  2 \left(2-4^{1-m} \right) (2m)! ~ \zeta(2 m) ,
\end{equation}
leading to a new expression for $I_1$: 
\begin{align}
I_1 =-T\, F\left(\frac{V}{2}  \right) -\sum_{m=1}^{\infty} & T^{2m+1} F^{(2m)}\left(\frac{V}{2}  \right) \nonumber \\
& \times (2-4^{1-m}) \zeta(2m) ~,
\end{align}
where the coefficients converge quite rapidly towards $2$, as can be seen from the first few values shown in Tab.~\ref{tab:coeff}.

\begin{table}
\caption{Coefficients entering the expansion of $I_1$.}
\renewcommand*{\arraystretch}{2.0}
\begin{tabular}{|>{$}c<{$}|>{$}c<{$}|>{$}c<{$}|>{$}c<{$}|>{$}c<{$}|>{$}c<{$}|}
\hline
m  & 1 & 2 & 3 & 4 & 5\\
\hline
\displaystyle \left( 2-4^{1-m} \right) \zeta(2m) &  \displaystyle \frac{\pi^2}{6} &  \displaystyle \frac{7 \pi^4}{360} & \displaystyle \frac{31 \pi^6}{15120} & \displaystyle \frac{127 \pi^8}{604800}& \displaystyle \frac{73 \pi^{10}}{3241440}\\[0.3em]
 \hline
\approx &  1.64493 & 1.89407 &  1.9711 &  1.99247 &  1.99808 \\
 \hline
\end{tabular}
\label{tab:coeff}
\end{table}

We now turn to the computation of $I_2$: 
\begin{align}
I_2 =&  \int_{-\infty}^{\infty} d\omega F\left(\omega +\frac{V}{2}  \right) \, g\left( \omega \right) \nonumber \\
=&  \int_{0}^{\infty} d\omega \left[ F\left(\frac{V}{2} -\omega  \right) + F\left(\frac{V}{2} +\omega  \right) \right]  f\left( \omega \right) \nonumber \\
& + \int_{0}^{-V} d\omega F\left(\omega +\frac{V}{2}  \right)  \tanh \left( \frac{\omega}{2T} \right) \, .
\end{align}
 Again using the Taylor expansion of $F$, we are left with an expression involving two sets of integrals, namely
\begin{align}
I_2 =& 2 \sum_{m=0}^{\infty} F^{(2m)} \left(\frac{V}{2}  \right) T^{2m+1} a_m  \nonumber \\
& + \sum_{k=0}^{\infty}  F^{(k)} \left(\frac{V}{2}  \right) T^{k+1} c_k \left(\frac{V}{T}\right) ~,
\label{eq:fullI2}
\end{align}
 with
  \begin{align}
  \label{eq:am}
 a_m &=  \frac{1}{(2m)!} \int_{0}^{\infty} d\epsilon \frac{\epsilon^{2m}}{1+e^\epsilon} ~,  \\
 c_k \left(\mu \right)  &=  \frac{1}{k !} \int_0^{-\mu} d\epsilon  \epsilon^k  \tanh \left( \frac{\epsilon}{2} \right) ~.
  \end{align}
   
The first set of integrals can be readily obtained as (see Ref.~\onlinecite{gradshteyn14}, 3.552.3)
 \begin{equation}
 a_m = \left\{ \begin{array}{rcl} 
 (1 - 4^{-m})  \zeta(2m+1) & : & m>0 \\
\ln(2) & : & m=0 
 \end{array}\right.
 \end{equation} 
 The remaining set of integrals can be obtained through a power expansion in its argument $\mu$ yielding
  \begin{equation}
 c_k \left(\mu \right)  =   \sum_{n=0}^{\infty} \frac{\mu^n}{n !} c_k^{(n)} (0) \, .
   \end{equation}
One readily sees that the first two terms of the expansion identically vanish for all values of $k$, while for $n \geq 2$, one has
\begin{align}
c_k^{(n)} (0) &= \frac{(-1)^n}{k !} \frac{d^{n-1}}{d \epsilon^{n-1}} \left[ \epsilon^k  \tanh \left( \frac{\epsilon}{2} \right) \right]_{\epsilon = 0} \nonumber \\
&=
\left\{ \begin{array}{ll} 
\frac{(-1)^n (n-1)!}{k !} {\cal C} \left(\frac{n-k}{2} \right) & \text{if $(n-k)>0$ and even} \\
0 & \text{otherwise}
 \end{array}\right.
   \end{align}
 where
 \begin{equation}
 {\cal C}(m) = 4 (-1)^{1+m} \frac{2^{2m}-1}{(2 \pi)^{2m}} \zeta(2m) ~.
 \label{eq:C2m}
   \end{equation}
The combination of Eqs.~\eqref{eq:am}-\eqref{eq:C2m} with Eq.~\eqref{eq:fullI2} leads to the full expression for $I_2$, which although complete is not totally practical. One option at this stage is to slightly simplify the resulting expression by resorting to a perturbative expansion in $V/T$. In particular, keeping all terms up to $O \left[ \left( \frac{V}{T} \right)^4 \right]$, this allows to write the set of integrals $ c_k \left(\frac{V}{T}\right)$ as
  \begin{align}
c_0 \left(\frac{V}{T}\right) =& \frac{1}{4} \left( \frac{V}{T} \right)^2 - \frac{1}{96} \left( \frac{V}{T} \right)^4 ~, \\
c_1 \left(\frac{V}{T}\right) =& - \frac{1}{4}  \left( \frac{V}{T} \right)^3  ~, \\
c_2 \left(\frac{V}{T}\right) =& \frac{1}{16}  \left( \frac{V}{T} \right)^4  ~.
  \end{align}
Note that all terms associated with $k \geq 3$ can be dropped as they only contribute to order $O \left[ \left( \frac{V}{T} \right)^5 \right]$ at best.

An alternative derivation involves writing the set of integrals in terms of the polylogarithm function $\text{Li}_n (z)$. In particular, one can readily show that
  \begin{align}
 c_k \left(\mu \right)  =&  \frac{\left( -\mu \right)^{k+1}}{(k+1) !}  - 2 \sum_{n=1}^{k+1} \frac{\left( -\mu \right)^{k+1-n}}{(k-n+1)!}  \text{Li}_n \left(-e^{\mu} \right) \nonumber \\
 & +  2 \text{Li}_{k+1} \left(-1 \right) ~,
  \end{align}
where one may bear in mind that there is a connection between the polylogarithm and the zeta functions, in particular $\text{Li}_{k+1} \left(-1 \right) = \left( 2^{-k} -1 \right) \zeta(k+1)$.

\bibliography{DeltaT_NDN_biblio}
\end{document}